\documentclass[english,10pt,aps,prd,a4paper,preprintnumbers,floatfix,nofootinbib,showpacs,superscriptaddress, notitlepage]{revtex4} 
 \pdfoutput=1
\usepackage[usenames,dvipsnames]{color}  
\usepackage{graphicx}
\usepackage{caption}
\usepackage{amsfonts}
\usepackage[export]{adjustbox} 
\usepackage{amsmath}
\usepackage{amssymb}
\usepackage{float}
\usepackage[colorlinks=true,citecolor=darkred,urlcolor=darkred, pdfborder={0 0 0}]{hyperref}
\usepackage[normalem]{ulem}
\usepackage{braket}

\makeatletter
\def\p@subsection{}
\makeatother

\definecolor{darkred}{rgb}{0.6,0,0}

\definecolor{linkcolor}{rgb}{0,0,0.5}

\def\gsim{\raise0.3ex\hbox{$\;>$\kern-0.75em\raise-1.1ex\hbox{$\sim\;$}}}
\def\lsim{\raise0.3ex\hbox{$\;<$\kern-0.75em\raise-1.1ex\hbox{$\sim\;$}}}

\def\beqn#1{\begin{equation}\label{#1}}
\def\eeqn{\end{equation}}

\def\beqa#1{\begin{eqnarray}\label{#1}}
\def\eeqa{\end{eqnarray}}

\def\0nbb {$0\nu\beta\beta$ }

\def\Z2{$\mathcal{Z_2}$}

\newcommand {\ignore}[1]{}

\newcommand{\sm}{{Standard Model }}

\def\gauge{$\mathrm{SU(3)_c \otimes SU(2)_L \otimes U(1)_Y \otimes U(1)_X\otimes \mathbb{Z}_2\otimes \mathbb{Z}'_2}$ }

\def\321{$\mathrm{SU(3) \otimes SU(2) \otimes U(1)}$ }

\newcommand{\AddrHBNI}{
	Homi Bhabha National Institute, BARC Training School Complex, Anushakti Nagar, Mumbai 400094, India }

\begin{document}

\bibliographystyle{unsrt} 

\title{Two-component scalar and fermionic dark matter candidates in a generic U$(1)_X$ model}
\author{Arindam Das}
\email{adas@particle.sci.hokudai.ac.jp}
\affiliation{Institute for the Advancement of Higher Education, Hokkaido University, Sapporo 060-0817, Japan}
\affiliation{Department of Physics, Hokkaido University, Sapporo 060-0810, Japan}
\author{Shivam Gola}
\email{shivamg@imsc.res.in}
\affiliation{The Institute of Mathematical Sciences,
C.I.T Campus, Taramani, Chennai 600 113, India}
\affiliation{\AddrHBNI}
\author{Sanjoy Mandal}
\email{smandal@kias.re.kr}
\affiliation{Korea Institute for Advanced Study, Seoul 02455, Korea}
\author{Nita Sinha}
\email{nita@imsc.res.in}
\affiliation{The Institute of Mathematical Sciences,
C.I.T Campus, Taramani, Chennai 600 113, India}
\affiliation{\AddrHBNI}
\preprint{EPHOU-21-017}
\begin{abstract}
We consider a $U(1)_X\otimes \mathbb{Z}_2\otimes \mathbb{Z}'_2$ extension of the Standard Model (SM), where the $U(1)_X$ charge of an SM field is given by a linear combination of its hypercharge and B$-$L  number. Apart from the SM particle content, the model contains three right-handed neutrinos (RHNs) $N_R^i$ and two scalars $\Phi$, $\chi$, all singlets under the SM gauge group but charged under $U(1)_X$ gauge group. Two of these additional fields, fermion $N_R^3$ is odd under $\mathbb{Z}_2$ and scalar $\chi$ is odd under $\mathbb{Z}'_2$ symmetry. Thus both $\chi$ and $N_R^3$ contribute to the observed dark matter relic density, leading to two-component dark matter candidates. We study in detail its dark matter properties such as relic density and direct detection taking into account the constraints coming from collider studies. We find that in our model, there can be possible annihilation of one Dark Matter (DM) into the other, which may potentially alter the relic density in a significant way.
\end{abstract}
\maketitle
\section{Introduction}
The two incontrovertible evidences of new physics (NP) are the existence of neutrino mass and DM. The former follows from the neutrino oscillation experiments~\cite{ParticleDataGroup:2020ssz} whereas the later is inferred from many observations such as galaxy rotation curves, galaxy clusters and large-scale cosmological data~\cite{Bertone:2004pz}. Underpinning the origin of neutrino mass and elucidating the nature of DM would constitute a major step forward in particle physics. Several simple extensions of the SM that can account for the DM have already been studied \cite{Jungman:1995df,Bertone:2004pz,Hooper:2007qk,McDonald:1993ex,Burgess:2000yq,LopezHonorez:2006gr,Barbieri:2006dq,Lopez-Honorez:2012tov}. In these models, the SM particle content is extended by additional fields, and a discrete symmetry is usually introduced to guarantee the stability of the DM particle in cosmological scale. In recent years, a class of models are proposed to incorporate the neutrino mass generation and the existence of DM in a unified framework. Motivated by this, people have studied well motivated beyond standard model (BSM) framework based on the gauged $U(1)_X$ model \cite{Okada:2016tci,Bandyopadhyay:2017bgh,Das:2019pua}. The most intriguing aspect of this model is that including three generations of right-handed neutrinos, as in the type-I seesaw process for creating light neutrino masses, is no longer an option, but emerges as the simplest solution to eliminate the gauge and mixed gauge-gravity anomalies \cite{Das:2016zue}. The scalar DM can be inherently stable in such models due to its $U(1)_X$ charge, but the fermionic DM cannot be realized in the simplest $U(1)_X$ model. Additional discrete symmetries can be introduced, which can stabilize one of the right-handed neutrinos to play the role of DM, while the other two neutrinos participate in the type I seesaw process to generate the required light neutrino masses and flavor mixing. Also, there are many models proposed in the literature, where neutrino mass generation is intimately connected with DM \cite{Ma:2006km,Hirsch:2013ola,Merle:2016scw,Avila:2019hhv,Mandal:2021yph,Mandal:2019oth}. In these type of models, DM is a mediator of neutrino mass generation.

Even though it is commonly considered that the observed relic density of DM can be explained entirely by a single particle, this is not always the case. Multi-component DM refers to a situation in which two or more particles contribute to the measured DM density. This has been already studied in many BSM scenarios \cite{Profumo:2009tb,Feldman:2010wy,Baer:2011hx,Aoki:2012ub,Bhattacharya:2013hva,Bian:2013wna,Kajiyama:2013rla,Esch:2014jpa,Bhattacharya:2016ysw,Bhattacharya:2019fgs,Bhattacharya:2019tqq,Bhattacharya:2017fid,Choi:2021yps,DiazSaez:2021pmg,Mohamadnejad:2021tke,Ho:2022erb}. In this article we consider a generic $U(1)_X$ model for two-component DM and discuss its phenomenological implications in this study. The general charge assignment of the particles in the model is obtained after the gauge and mixed gauge gravity anomalies. It further allows to study the dependence of the DM relic abundance on these charges which will appear at different interactions between the potential DM candidates, mediators and the other particles present in the model. In the context of the two-component DM system such properties have not been discussed in the previous literatures. Among these two DM candidates, one is scalar $\chi$ and other one is fermion $N_R^3$. $N_R^3$ and $\chi$ are SM singlets but charged under $U(1)_X$. $N_R^3$ is odd under a discrete symmetry $\mathbb{Z}_2$ and $\chi$ is odd under another discrete symmetry $\mathbb{Z}_2'$. The model is not exactly same as the singlet fermionic \cite{Lopez-Honorez:2012tov} plus scalar \cite{Burgess:2000yq} DM due to the presence of new gauge boson $Z'$ which couples with $N_R^3$ or $\chi$ as both are charged under $U(1)_X$. Hence, there will be many new processes which contribute to the relic density. Also, this model contains additional features such as annihilation of one type of DM into the other, which we investigate in some depth. We further examine the direct detection prospects of this model. We find that there is parameter space where both these particles can produce observable signals in future direct detection experiments. In this analysis we employ latest bounds on the $U(1)_X$ coupling from $Z^\prime$ mediated dilepton and dijet scenarios for different combinations of $U(1)_X$ charges. We estimate the results taking the allowed couplings after considering a variety of experimental searches. In addition to that we consider the mixing between the SM Higgs doublet and SM singlet $U(1)_X$ scalar. Considering the allowed parameter of this mixing angle we estimate the relic abundance of a viable DM candidate in the model. 

The paper is organized as follows. In sec.~\ref{sec:model} we introduce the model and discuss the details of the new fields and their interactions. In sec.~\ref{sec:constraints} we discuss different theoretical and experimental constraints on the model parameters. In sec.~\ref{sec:dark-matter} we discuss in detail the relic density and direct detection properties coming from our two-component DM candidates. In sec.~\ref{sec:relic-dependence-xH} we study the relic density dependence on $U(1)_X$ charge in the case of $Z'$-portal DM. Finally in sec.~\ref{sec:conclusion} we conclude the article.
\begin{table}[t]
\begin{center}
\begin{tabular}{||c|ccc||c||c|c||}
\hline
\hline
            & SU(3)$_c$ & SU(2)$_L$ & U(1)$_Y$ & U(1)$_X$  & $\mathbb{Z}_2$  & $\mathbb{Z}'_2$\\[2pt]
\hline
\hline
&&&&&&\\[-12pt]
$q_L^i$    & {\bf 3}   & {\bf 2}& $\frac{1}{6}$ & 		 $x_q= \frac{1}{6}x_H + \frac{1}{3}x_\Phi$  & + & +  \\[2pt] 
$u_R^i$    & {\bf 3} & {\bf 1}& $\frac{2}{3}$ &  	  $x_u= \frac{2}{3}x_H + \frac{1}{3}x_\Phi$ & + & + \\[2pt] 
$d_R^i$    & {\bf 3} & {\bf 1}& $-\frac{1}{3}$ & 	 $x_d=-\frac{1}{3}x_H + \frac{1}{3}x_\Phi$ & + & + \\[2pt] 
\hline
\hline
&&&&&&\\[-12pt]
$\ell_L^i$    & {\bf 1} & {\bf 2}& $-\frac{1}{2}$ & 	 $x_\ell =- \frac{1}{2}x_H - x_\Phi$ & + & +  \\[2pt] 
$e_R^i$   & {\bf 1} & {\bf 1}& $-1$   &		$x_e= - x_H - x_\Phi$ & + & +  \\[2pt] 
\hline
\hline
$N_R^{1,2}$   & {\bf 1} & {\bf 1}& $0$   &	 $x_\nu =- x_\Phi$ & + & + \\[2pt] 
$N_R^{3}$   & {\bf 1} & {\bf 1}& $0$   & 	 $x_\nu=- x_\Phi$  & - & + \\[2pt] 
\hline
\hline
&&&&&&\\[-12pt]
$H$         & {\bf 1} & {\bf 2}& $\frac{1}{2}$  &  	 $\frac{x_H}{2}$ & + & + \\ 
$\Phi$      & {\bf 1} & {\bf 1}& $0$  & 	 $2 x_\Phi$  & +  &  +  \\ 
$\chi$      & {\bf 1} & {\bf 1}& $0$  & 	 $-x_\Phi$  & +  & -  \\ 
\hline
\hline
\end{tabular}
\end{center}
\caption{
Particle content of  the minimal U$(1)_X$ model where $i(=1, 2, 3)$ represents the family index. The scalar charges $x_H$, $x_\Phi$ are the real parameters. The U$(1)_X$ gauge coupling is a free parameter in this model.}
\label{tab1}
\end{table}
\section{Model}
\label{sec:model}
The considered model is a general but minimal U$(1)_X$ extension of the SM where in addition to the SM particles, three generations of right handed neutrinos~($N_R^i$) and two U$(1)_X$ complex scalar fields~($\Phi,\chi$) are included. The SM as well as new particles and their charges are given in Table~\ref{tab1}, where the family index $i$ runs from 1 to 3. The general charge assignment can be reduced to a special and unique form after cancelling the gauge and mixed gauge-gravity anomalies. To do that we describe the Yukawa, scalar and gauge sectors of the model in the following:
\subsection{Yukawa Sector}
The Yukawa sector of the model can be written in a gauge-invariant way as
\begin{align}
 \mathcal{L}_{y} & =-\sum_{i,j=1}^3 y_{u}^{ij}\overline{q_{L}^{i}}\tilde{H}u^{j}_{R} - \sum_{i,j=1}^3 y_{d}^{ij}\overline{q_{L}^{i}}H d^{j}_{R} - \sum_{i,j=1}^3 y_{e}^{ij}\overline{\ell_{L}^{i}} H e^{j}_{R}- \sum_{i=1}^3\sum_{j=1}^2 y_{\nu}^{ij}\overline{\ell_{L}^{i}}\tilde{H} N^{j}_{R} \nonumber \\
& -\frac{1}{2}\sum_{i,j=1}^2 y_{M}^{ij}\Phi\overline{N_{R}^{ic}} N^{j}_{R}
  -\frac{1}{2} y_{M}^{3}\Phi\overline{N_{R}^{3c}}N^{3}_{R} +\text{H.c.}
\label{Yukawa}
\end{align}
The $U(1)_X$ charges of the particles are controlled by two parameters only, $x_H$ and $x_\Phi$, as seen in Table~\ref{tab1} and can be defined as a linear combination of the SM $U(1)_Y$ and the $U(1)_{B - L}$ which can be obtained after solving the gauge and mixed gauge-gravity anomaly cancellation equations. The detailed equations are given in Appendix.~\ref{anomaly} from \cite{Das:2016zue}. Note that the $B-L$ case can be obtained with the choice $x_H=0$ and $x_\Phi=1$. For simplicity we fix $x_\Phi=1$ in our analysis throughout the paper. In addition to that we find if $x_H=-2$, then the left handed fermions have no interactions with the $Z^\prime$ leading to an $U(1)_R$ scenario. For $x_H=-1$ and $1$ the interactions of $d_R$ and $e_R$ with $Z^\prime$ are switched off. The model is constructed such that all SM fields are even under the discrete symmetry $\mathbb{Z}_2\otimes \mathbb{Z}'_2$, where as $N_R^3$ is odd under $\mathbb{Z}_2$ and $\chi$ is odd under $\mathbb{Z}'_2$. The DM interactions where $Z^\prime$ participates in any of the vertices will manifest the dependence of $x_H$ and $x_\Phi$ respectively.
Last two terms will give the Dirac and Majorana contributions to the neutrino mass generation. We have assumed a basis in which $y_M^{ij}$ is diagonal $y_M=\text{diag}(y_M^1,y_M^2)$, without the loss of generality. Relevant light neutrino masses will come from the fourth and fifth term of Eq.~\ref{Yukawa}. After the electroweak symmetry breaking we can write the mass terms as,
 \begin{align}
  -\mathcal{L}_{M}=\sum_{j=1}^3\sum_{k=1}^2\overline{\nu_{jL}}m^{jk}_{D} N_{kR}+\frac{1}{2}\sum_{j,k=1}^2\overline{(N_{R})_{j}^{c}}M_R^{jk} N_{kR}+\text{H.c.},
 \end{align}
where $m_D^{jk}=\frac{y_{\nu}^{jk}v_H}{\sqrt{2}}$ and $M_R^{jk}=\frac{y_{M}^{jk}}{\sqrt{2}}v_\Phi$. Now we can write the $\mathcal{L}_{M}$ in the following matrix form,
\begin{align}
 -\mathcal{L}_{M}=\frac{1}{2}
 \begin{pmatrix}
  \overline{\nu_{L}} & \overline{(N^1_{R})^{c}} & \overline{(N^2_{R})^{c}} \\
 \end{pmatrix}
 \begin{pmatrix}
  0_{3\times 3}   &  (m_{D})_{3\times 2} \\
  (m_{D}^T)_{2\times 3}  &  (M_{R})_{2\times 2} \\
 \end{pmatrix}
\begin{pmatrix}
 (\nu_{L})^{c}\\
 N^1_{R} \\
 N^2_{R} \\
\end{pmatrix}
\end{align}
From this mass matrix, using the assumption $m_{D}\ll M_{R}$, it is easy to recover the seesaw formula for the light Majorana neutrinos as, $\mathcal{M}_\nu \approx m_D M_R^{-1}m_D^T$ and the heavy neutrino mass as, $M_{N}\approx M_{R}$. We emphasize that the right-handed neutrino $N_R^3$ is decoupled by construction from this seesaw mechanism as it is odd under the $\mathbb{Z}_2$ symmetry.
\subsection{Scalar Sector}
We begin by writing down the Lagrangian of the scalar sector. Apart from the SM Higgs doublet $H$ we have two complex scalars $\Phi$ and $\chi$, both charged under $U(1)_X$, but with $\mathbb{Z}_2$ even parity. The most general renormalizable and \gauge gauge invariant scalar sector can be written as
\begin{align}
 \mathcal{L}_{s}=(D^{\mu}H)^{\dagger}(D_{\mu}H)+(D^{\mu}\Phi)^{\dagger}(D_{\mu}\Phi)+(D^{\mu}\chi)^{\dagger}(D_{\mu}\chi)-V(H,\Phi,\chi),
\end{align}
where the covariant derivative is defined as, $D_{\mu}=\partial_{\mu}-ig_{s}T^{a}G^{a}_{\mu}-igT^{a}W^{a}_{\mu}-ig_{1}YB_{\mu}-ig_{1}^{'}Y_{X}B_{\mu}^{'}$. $B_{\mu}^{'}$
is the $U(1)_{X}$ gauge fields. The $U(1)_X$ gauge coupling $g_{1}^{'}$ is a free parameter. The scalar potential $V(H,\Phi,\chi)$ is given by,
\begin{align}
 V(H,\Phi,\chi) = &-\mu_{H}^{2}H^{\dagger}H-\mu_{\Phi}^{2}\Phi^{\dagger}\Phi + m_{\chi}^{2}\chi^{\dagger}\chi +\lambda_{H}(H^{\dagger}H)^{2}+\lambda_{\Phi}(\Phi^{\dagger}\Phi)^{2}
 +\lambda_{\chi}(\chi^{\dagger}\chi)^{2}\nonumber \\
 &+\lambda_{H\Phi}(H^{\dagger}H)(\Phi^{\dagger}\Phi)\nonumber 
 +\lambda_{\Phi\chi}(\Phi^{\dagger}\Phi)(\chi^{\dagger}\chi)+\lambda_{H\chi}(H^{\dagger}H)(\chi^{\dagger}\chi)+(\lambda_{\Phi\chi\chi}\Phi\chi\chi + \text{H.c.})
\end{align}
The breaking of the electroweak and the $U(1)_X$ gauge symmetries are driven by the vacuum expectation values~(vev) of the scalar fields $H$ and $\Phi$ as the field $\chi$ does not get any vev due to $\mathbb{Z}'_2$ symmetry protection. Denoting the vevs of $H$ and $\Phi$ by $v_H$ and $v_\Phi$, the fields $H,\Phi$ and $\chi$ can be written in unitary gauge after symmetry breaking in the form
\begin{align}
 H=\frac{1}{\sqrt{2}}
 \begin{pmatrix}
  \phi^+ \\
  v_H+R_1+i I_1  \\
 \end{pmatrix},\hspace{0.2cm}
 \Phi=\frac{1}{\sqrt{2}}(v_\Phi+R_2+i I_2),
 \hspace{0.2cm}
 \chi=\frac{1}{\sqrt{2}}(\chi_{R}+i\chi_{I})
\end{align}
$\Phi^{\pm}$ are the would be Goldstone boson of $W^{\pm}$, while $I_1$ and $I_2$ will mix to give the Goldstone bosons of the $Z$ and $Z^{'}$ bosons, respectively. The mass matrix of CP-even Higgs scalars in the basis $(R_1 , R_2 )$ reads as
\begin{align}
M_R^2=\begin{bmatrix}
2\lambda_H v_H^2 &  \lambda_{H\Phi} v_H  v_\Phi \\
\lambda_{H\Phi} v_H v_\Phi  &  2\lambda_{\Phi}v_\Phi^2 \\
\end{bmatrix}
\end{align}
with the mass eigenvalues given by
\begin{align}
m_{h_{1}}^{2}=\lambda_{H}v_H^{2}+\lambda_{\Phi}v_\Phi^{2}-\sqrt{(\lambda_{H}v_H^{2}-\lambda_{\Phi}v_\Phi^{2})^{2}+(\lambda_{H\Phi}v_H v_\Phi)^{2}} \\
m_{h_{2}}^{2}=\lambda_{H}v_H^{2}+\lambda_{\Phi}v_\Phi^{2}+\sqrt{(\lambda_{H}v_H^{2}-\lambda_{\Phi}v_\Phi^{2})^{2}+(\lambda_{H\Phi}v_H v_\Phi)^{2}}
\end{align}
where the scalars $h_1$ and $h_2$ have masses $m_{h_1}$ and $m_{h_2}$ respectively, and by convention $m_{h_1}^2 \leq m_{h_2}^2$ throughout this work. We have identified $h_1$ as the SM Higgs discovered at LHC, with mass $m_{h_1}=125$~GeV. The two mass eigenstates $h_i$ are related with the $(R_1, R_2)$ fields through the rotation matrix $O_R$ as,
\begin{align}
\begin{bmatrix}
h_1 \\
h_2 \\
\end{bmatrix}=O_R
\begin{bmatrix}
R_1 \\
R_2 \\
\end{bmatrix}=\begin{bmatrix} \cos\alpha  &  \sin\alpha \\
-\sin\alpha  &  \cos\alpha  \\  \end{bmatrix}
\begin{bmatrix}
R_1 \\
R_2 \\
\end{bmatrix},
\label{eq:rotation}
\end{align}
where $\alpha$ is the mixing angle. The rotation matrix satisfies
\begin{align}
O_R M_R^2 O_R^T=\text{diag}\left(m_{h_1}^2,m_{h_2}^2\right).
\label{eq:diag}
\end{align}
We can use Eq.~\eqref{eq:rotation} and \eqref{eq:diag} to solve for the potential parameters $\lambda_H$, $\lambda_\Phi$ and $\lambda_{H\Phi}$ in terms of the mixing angle $\alpha$ and the scalar masses $m_{h_i}$ as
\begin{align}
 &\lambda_{H}=\frac{m_{h_{2}}^{2}}{4v_H^{2}}(1-\text{cos}~2\alpha)+\frac{m_{h_{1}}^{2}}{4v_H^{2}}(1+\text{cos}~2\alpha)\\
 & \lambda_{\Phi}=\frac{m_{h_{1}}^{2}}{4v_\Phi^{2}}(1-\text{cos}~2\alpha)+\frac{m_{h_{2}}^{2}}{4v_\Phi^{2}}(1+\text{cos}~2\alpha)\\
 & \lambda_{H\Phi}=\text{sin}~2\alpha\left(\frac{m_{h_{1}}^{2}-m_{h_{2}}^{2}}{2v_\Phi v_H}\right)
 \label{coup}
\end{align}
Exact conservation of the $\mathbb{Z}'_2$ symmetry forbids the mixing between $\chi$-Higgs and $\chi$-$\Phi$. The real and imaginary components of $\chi$ have the following masses
\begin{align}
 M_{\chi_R}^2=m_{\chi}^{2}+v_\Phi^{2}\frac{\lambda_{\Phi\chi}}{2}+v_H^{2}\frac{\lambda_{H\chi}}{2}+\sqrt{2}v_\Phi\lambda_{\Phi\chi\chi}\\
M_{\chi_I}^2=m_{\chi}^{2}+v_\Phi^{2}\frac{\lambda_{\Phi\chi}}{2}+v_H^{2}\frac{\lambda_{H\chi}}{2}-\sqrt{2}v_\Phi\lambda_{\Phi\chi\chi}
\end{align}
The difference $M_{\chi_R}^2-M_{\chi_I}^2$ depends only on the parameter $\lambda_{\Phi\chi\chi}$. The conservation of the $\mathbb{Z}'_2$ symmetry also makes the lightest of the two eigenstates $\chi_R$ and $\chi_I$ a viable scalar DM candidate.
\subsection{Gauge sector}
To determine the gauge boson spectrum, we have to expand the scalar kinetic terms and replace 
\begin{align}  
H=\frac{1}{\sqrt{2}}
\begin{pmatrix}
0   \\
v_H+R_1  \\
\end{pmatrix},\,\,\,\,\text{and  }\,\Phi=\frac{v_\Phi + R_2}{\sqrt{2}},
\end{align}
With this above replacement we can expand the scalar kinetic terms $(D^{\mu}H)^{\dagger}(D_{\mu}H)$ and $(D^{\mu}\Phi)^{\dagger}(D_{\mu}\Phi)$ as follows
\begin{align}
 (D^{\mu}H)^{\dagger}(D_{\mu}H)&\equiv\frac{1}{2}\partial^{\mu}R_1\partial_{\mu}R_1+\frac{1}{8}(R_1+v_H)^{2}\Big(g^{2}|W_{1}^{\mu}-iW_{2}^{\mu}|^{2}+(gW_{3}^{\mu}-g_{1}B^{\mu}-\tilde{g}B^{'\mu})^{2}\Big)\\
 (D^{\mu}\Phi)^{\dagger}(D_{\mu}\Phi)&\equiv\frac{1}{2}\partial^{\mu}R_2\partial_{\mu}R_2+\frac{1}{2}(R_2+v_\Phi)^{2}(2g_{1}^{''} B^{'\mu})^{2}.
\end{align}
where we have defined $\tilde{g}=g_{1}^{'}x_H$ and $g_{1}^{''}=g_{1}^{'}x_\Phi$. SM charged gauge boson $W^{\pm}$ can be easily recognised with mass $M_{W}=\frac{gv_H}{2}$. Linear combination of $B^{\mu}$, $W_{3}^{\mu}$ and $B^{'\mu}$ gives definite mass eigenstates $A^{\mu}$, $Z^{\mu}$ and $Z^{'\mu}$,
\begin{align}
 \begin{pmatrix}
  B^{\mu}\\
  W_{3}^{\mu}\\
  B^{'\mu}
 \end{pmatrix}
=
\begin{pmatrix}
 \text{cos}~\theta_{w}  & -\text{sin}~\theta_{w}~\text{cos}~\theta^{'}  &  \text{sin}~\theta_{w}~\text{sin}~\theta^{'} \\
 \text{sin}~\theta_{w}  & \text{cos}~\theta_{w}~\text{cos}~\theta^{'}  &  -\text{cos}~\theta_{w}~\text{sin}~\theta^{'} \\
 0                      & \text{sin}~\theta^{'}                        &  \text{cos}~\theta^{'} \\
\end{pmatrix}
\begin{pmatrix}
 A^{\mu}\\
 Z^{\mu} \\
 Z^{'\mu} \\
\end{pmatrix}
\end{align}
where $\theta_{w}$ is the Wienberg mixing angle and,
\begin{align}
 \text{tan}~2\theta^{'}=\frac{2\tilde{g}\sqrt{g^{2}+g_{1}^{2}}}{\tilde{g}^{2}+16\left(\frac{v_\Phi}{v_H}\right)^{2}g_{1}^{''2}-g^{2}-g_{1}^{2}}.
\end{align}
Masses of $A,\,Z$ and $Z^{'}$ are given by,
\begin{align}
 M_{A}=0,
 \hspace{0.5cm}
 M_{Z,Z^{'}}^{2}=\frac{1}{8}\left(Cv_H^{2}\mp\sqrt{-D+v_H^{4}C^{2}}\right),
\end{align}
where,
\begin{align}
 C=g^{2}+g_{1}^{2}+\tilde{g}^{2}+16\left(\frac{v_\Phi}{v_H}\right)^{2}g_{1}^{''2},
 \hspace{0.5cm}
 D=64v_H^{2}v_\Phi^{2}(g^{2}+g_{1}^{2})g_{1}^{''2}.
\end{align}
\section{Theoretical and experimental constraints}
\label{sec:constraints}
We discuss about different constraints on the model parameters such as $U(1)_X$ gauge coupling and scalar mixing angle. To estimate the constraints we consider vacuum stability, perturbative unitarity, collider searches of BSM Higgs and $Z^\prime$ boson respectively. 

\subsection{Vacuum Stability}
The above scalar potential must be bounded from below.  To determine the conditions for $V(H,\Phi,\chi)$ to be bounded from below, we need to check the following symmetric matrix which comes from the quadratic part of the potential,
\begin{align}
 V^4_s=
\begin{pmatrix}
 \lambda_{H}    &   \frac{\lambda_{H\Phi}}{2}    &   \frac{\lambda_{H\chi}}{2}  \\
 \frac{\lambda_{H\Phi}}{2}  & \lambda_{\Phi}  &    \frac{\lambda_{\Phi\chi}}{2}  \\
 \frac{\lambda_{H\chi}}{2}  &  \frac{\lambda_{\Phi\chi}}{2}  & \lambda_{\chi}  \\
\end{pmatrix}
\end{align}
Requiring such a matrix to be positive-definite, we obtain the following conditions,
\begin{align}
 &\lambda_{H} > 0,
 \hspace{1cm}
 4\lambda_{H}\lambda_{\Phi}-\lambda_{H\Phi}^{2}>0 ,\nonumber \\
 &(-\lambda_{H}\lambda_{\Phi\chi}^{2}+\lambda_{H\Phi}\lambda_{\phi\chi}\lambda_{H\chi}-\lambda_{\Phi}\lambda_{H\chi}^{2}+4\lambda_{H}\lambda_{\Phi}\lambda_{\chi}-\lambda_{H\Phi}^{2}\lambda_{\chi})>0.
\label{eq:stability}
\end{align}
To have an absolutely stable vacuum, one needs to satisfy the condition given in Eq.~\eqref{eq:stability} at each and every energy scale. To ensure perturbativity, we take a conservative approach of simply requiring that $\lambda_i\leq 4\pi$ and $g_i (x_H, x_\Phi) \leq \sqrt{4\pi}$ where $g_i$ denotes the respective gauge couplings of the model.
\subsection{Higgs Invisible decay}
\label{sec:inv-Higgs}
In this section, we discuss the constraints on the relevant parameter space of Higgs bosons which follow from searches performed at LHC. Before discussing the collider constraints, notice that due to the presence of this heavy Higgs $h_2$, the coupling of the \sm Higgs boson to \sm particles gets modified according to the substitution rule
\begin{align}
h_{\text{SM}}\to \cos\alpha \,h_1 - \sin\alpha\, h_2
\label{eq:substitution}
\end{align}
If either of $M_{\chi_R}$ or $M_{\chi_I}$ is smaller than half of the Higgs mass then these two channels will also contribute to the invisible mode. The partial decay width to $\chi_{R}\chi_R$ and $\chi_I\chi_I$ are given as follows:
\begin{align}
\Gamma(h_1\to\chi_R\chi_R)&=\frac{1}{32\pi m_{h_1}}\big(\lambda_{H\chi}v_H \cos\alpha + \lambda_{\Phi\chi}v_\Phi\sin\alpha + \sqrt{2}\lambda_{\Phi\chi\chi}\sin\alpha \big)^2\sqrt{1-\frac{4M_{\chi_R}^2}{m_{h_1}^2}}\\
\Gamma(h_1\to\chi_I\chi_I)&=\frac{1}{32\pi m_{h_1}}\big(\lambda_{H\chi}v_H \cos\alpha + \lambda_{\Phi\chi}v_\Phi\sin\alpha - \sqrt{2}\lambda_{\Phi\chi\chi}\sin\alpha \big)^2\sqrt{1-\frac{4M_{\chi_I}^2}{m_{h_1}^2}}
\end{align}
Hence the total invisible decay width of \sm Higgs boson $h_1$ is given as
\begin{align}
\Gamma^{\text{inv}}(h_1)=\Gamma(h_1\to\chi_R\chi_R)+\Gamma(h_1\to\chi_I\chi_I)
\end{align}
Accordingly the invisible branching ratio for $h_1$ is given by 
\begin{align}
\text{BR}^{\text{inv}}(h_1)=
\frac{\Gamma^{\text{inv}}(h_1)}{\cos^2\alpha\Gamma^\text{SM}(h_1)+\Gamma^{\text{inv}}(h_1)},
\label{eq:inv-BR-Higgs}
\end{align}
where $\Gamma^\text{SM}(h_1)=4.1$ MeV. The current upper bound on the branching ratio to invisible decay modes from CMS experiment~\cite{CMS:2018yfx}~\footnote{The present bound from ATLAS for invisible Higgs decays is $\text{BR}^{\text{inv}}(h_1) \leq 0.26$~\cite{ATLAS:2019cid}.},
\begin{align}
\text{BR}^{\text{inv}}(h_1) \leq 0.19.
\label{eq:invisible}
\end{align}
In the case of $M_\chi < m_{h_1}/2$ and $\sin\alpha\sim 0$, the invisible Higgs decay constraint can be translated as an upper bound on the quartic coupling $\lambda_{H\chi}$: 
\begin{align}
\lambda_{H\chi}\left(1-\frac{4M_\chi^2}{m_{h_1}^2}\right)^{\frac{1}{4}}\leq 9.8\times 10^{-3}.
\end{align}
Notice that in case of $m_{\chi}<m_h/2$ and non-zero values of mixing parameter $\sin\alpha$, the exclusion limit on $\lambda_{H\chi}$ depends on other quartic couplings such as $\lambda_{\Phi\chi}$, $\lambda_{\Phi\chi\chi}$ and on the vev $v_\Phi$.
\subsection{Collider constraints on $g_1'-M_{Z'}$}
\label{sec:collider-constraints}
\begin{figure}[h!]
\includegraphics[width=0.45\textwidth]{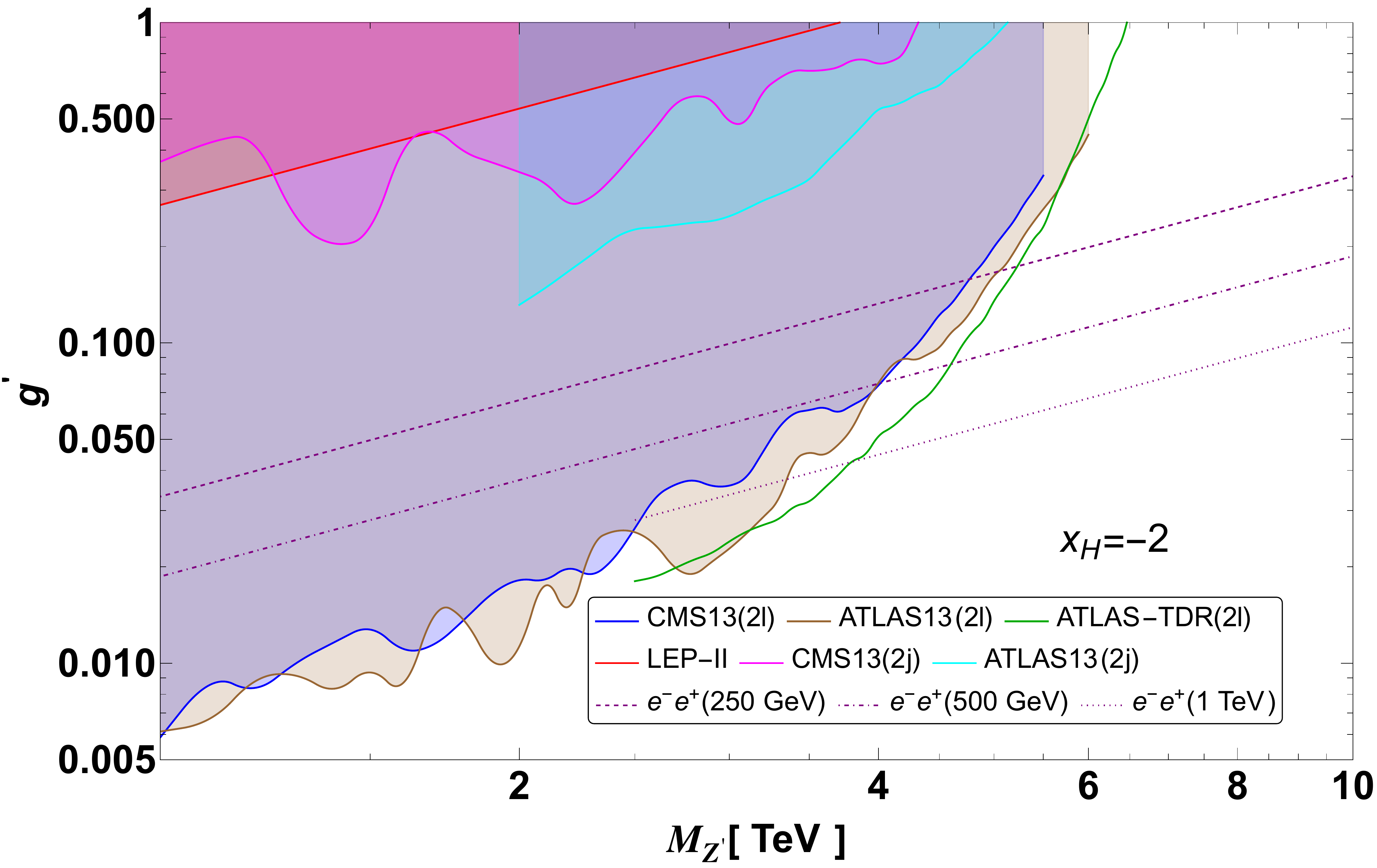}
\includegraphics[width=0.45\textwidth]{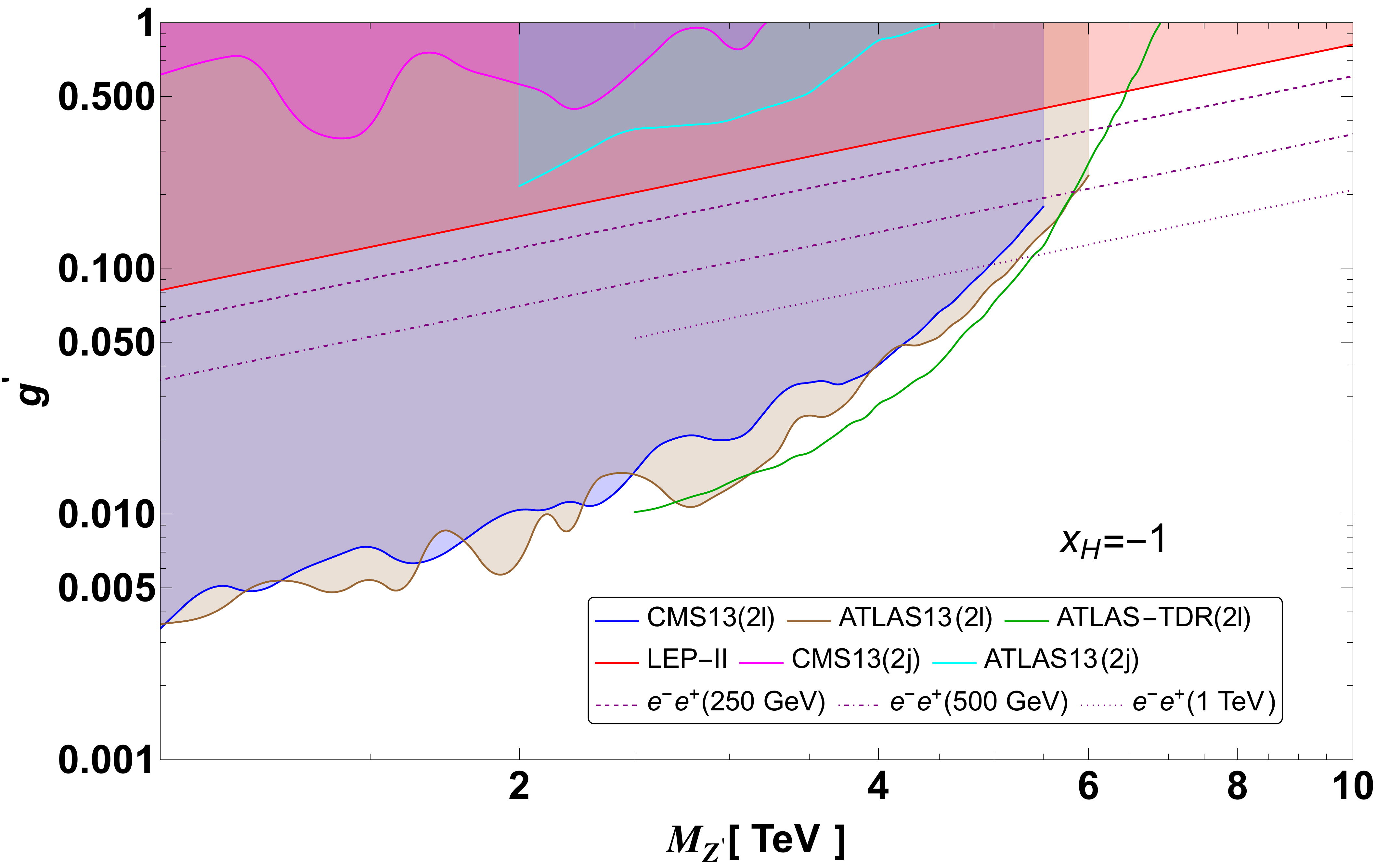}\\
\includegraphics[width=0.45\textwidth]{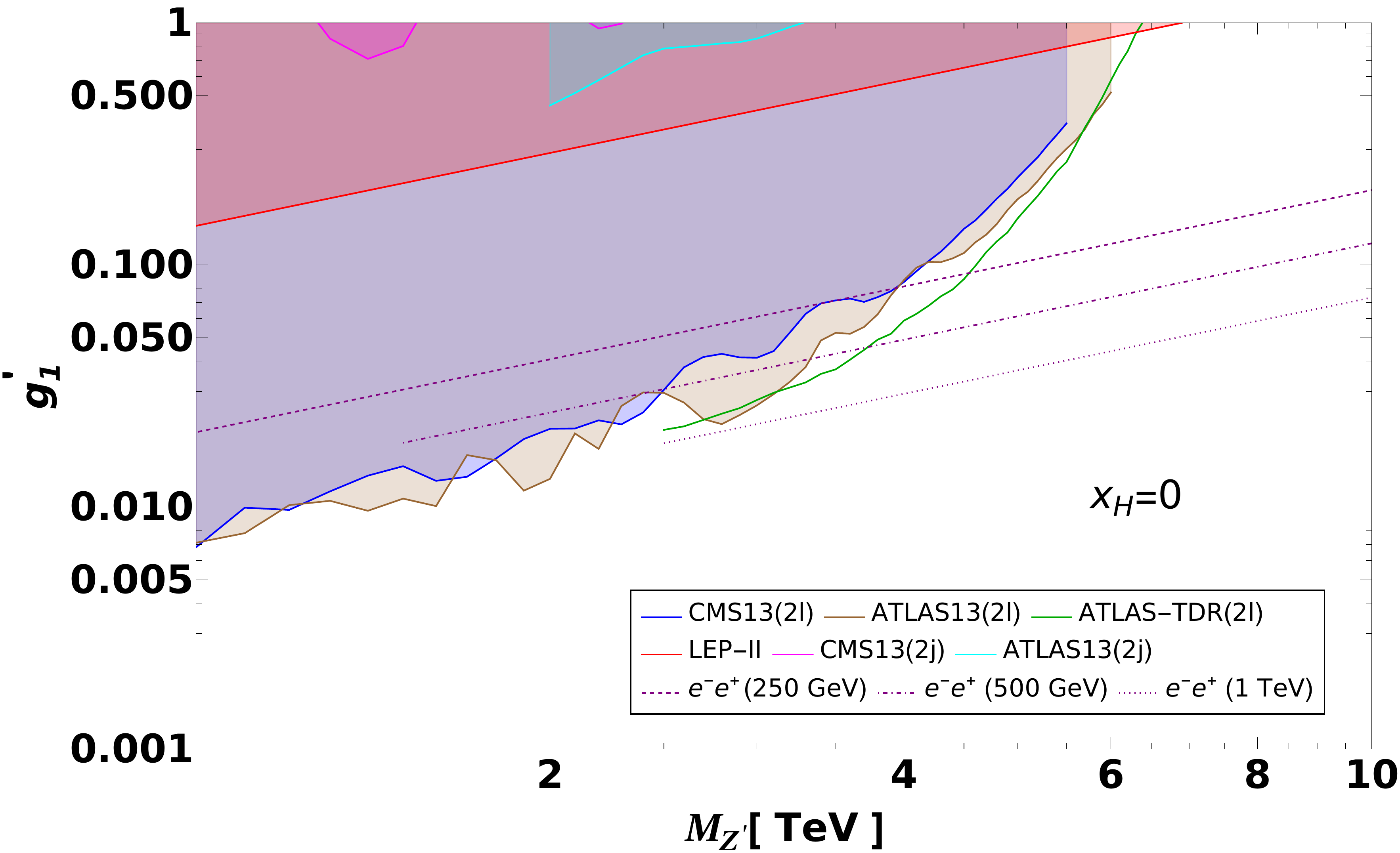}\\
\includegraphics[width=0.45\textwidth]{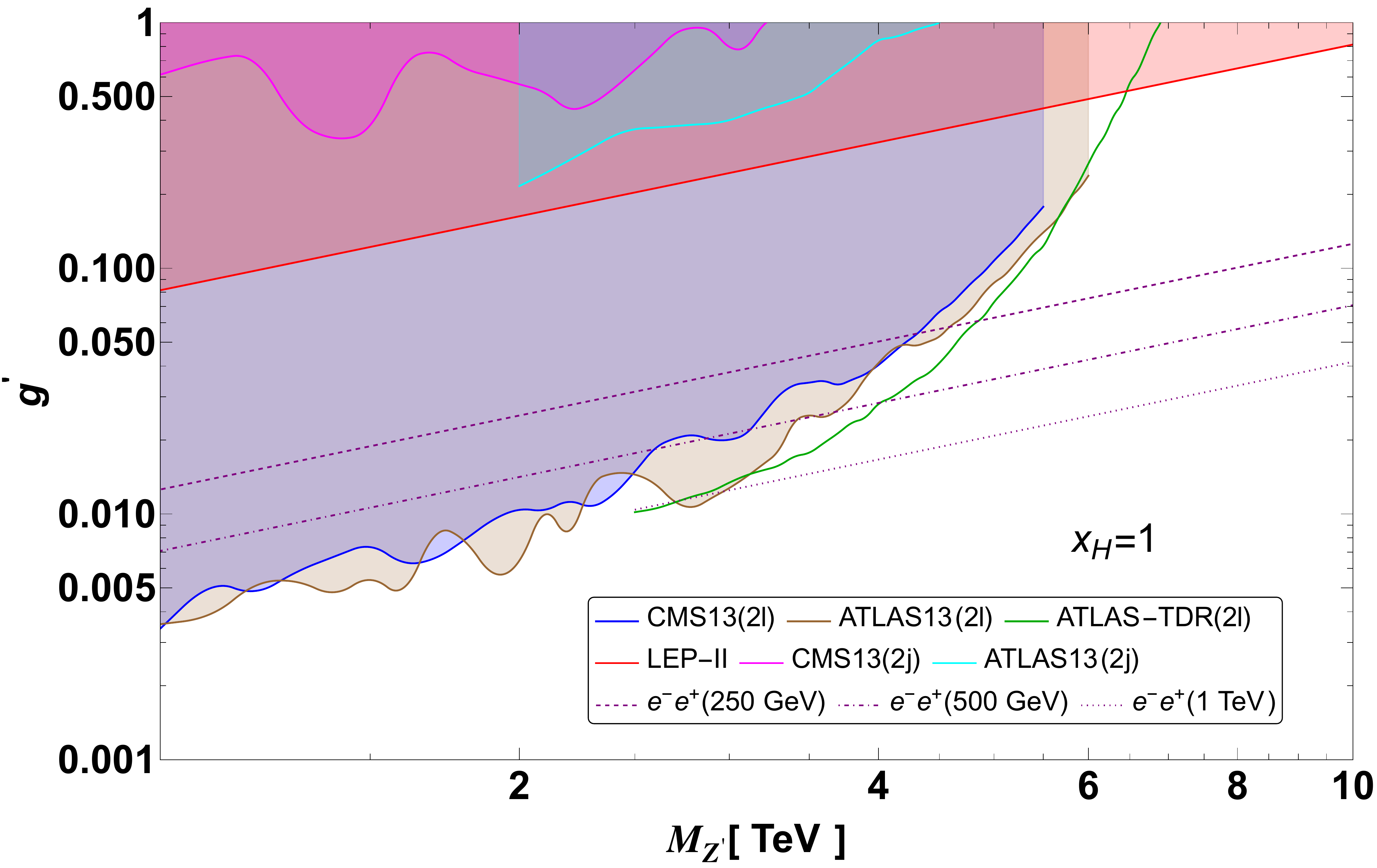}
\includegraphics[width=0.45\textwidth]{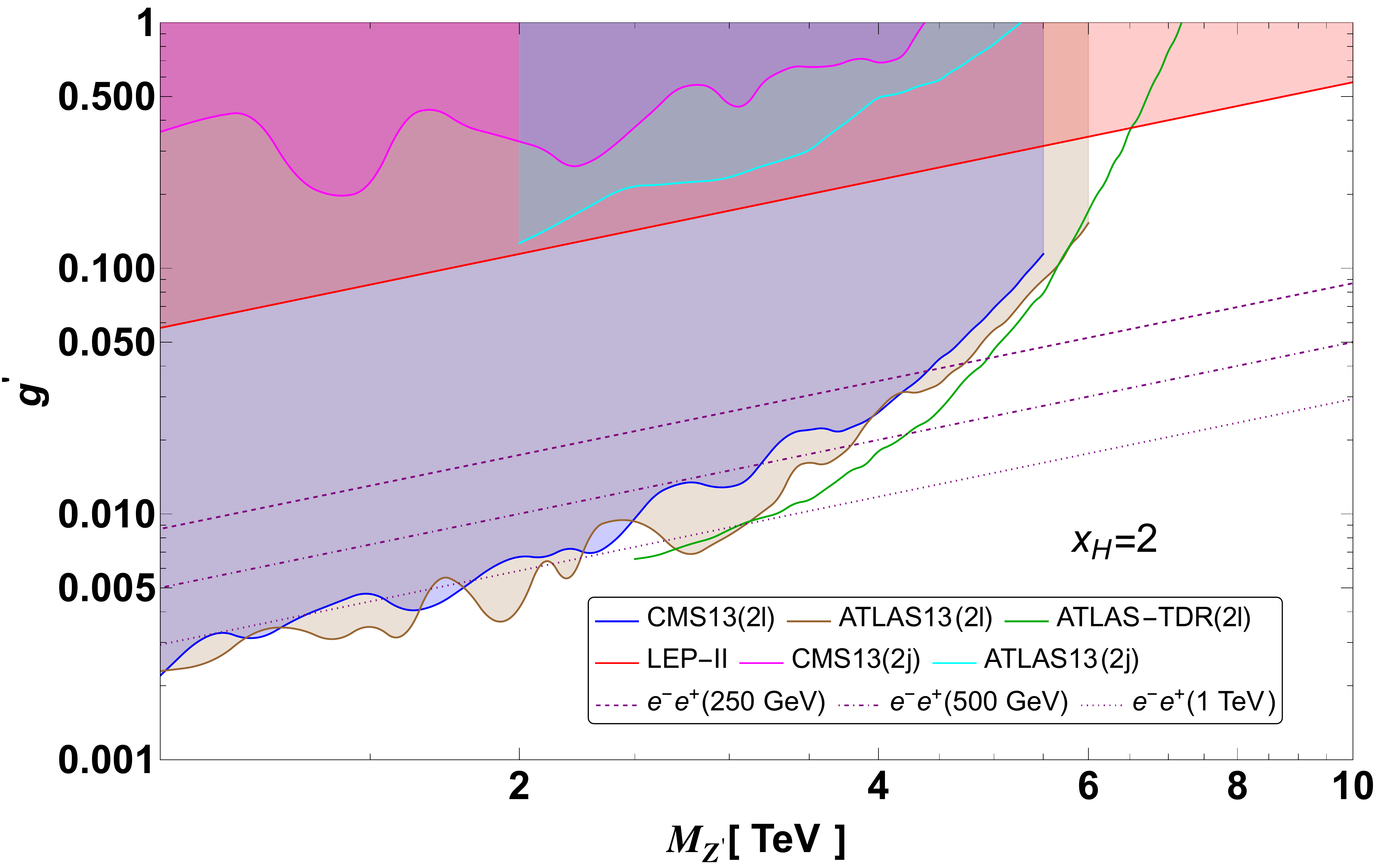}
\caption{Exclusion limits on $g_1'$ as a function of $M_{Z'}$ for various values of $x_H$ and with $x_{\Phi}=1$. The shaded regions are excluded by the current experimental data from LEP-II~\cite{ALEPH:2013dgf}, and LHC dilepton~\cite{ATLAS:2019erb,CMS:2019tbu}, LHC dijet~\cite{ATLAS:2019bov,CMS:2018mgb} searches. Also, we have shown in unshaded curves the prospected limit coming from future HL-LHC~\cite{CERN-LHCC-2017-018} as well as the ILC~\cite{Das:2021esm}.}
\label{fig:collider-constraints}
\end{figure}

In this section we will employ the most recent collider results to derive constraints on the model parameters such as $g_{1}', M_{Z'}$ and $x_H$. In the ATLAS and CMS collaborations analysis, sequential SM $Z'$ model~\cite{Barger:1980ti} has been considered as a reference model. We can easily translate the constraints of sequential SM $Z'$ model to our $U(1)_X$ model parameters. For example, we can obtain limits on $M_{Z'}/g_{1}'$ for different values of $x_H$ with fixed $x_{\Phi}=1$, as shown in Fig.~\ref{fig:collider-constraints}. The various shaded regions in Fig.~\ref{fig:collider-constraints} shows the excluded limit from ATLAS and CMS search for $Z'$ in both dilepton~\cite{ATLAS:2019erb,CMS:2019tbu} and dijet~\cite{ATLAS:2019bov,CMS:2018mgb} channels. We also consider the future high-luminosity phase of the 14 TeV LHC (HL-LHC) with 3 $\text{ab}^{-1}$ integrated luminosity and draw the projected dilepton bounds following the analysis
given in the ATLAS technical design report (TDR)~\cite{CERN-LHCC-2017-018}. The LEP-II~\cite{ALEPH:2013dgf} exclusion is shown by the red-shaded region, while the future ILC prospects~\cite{Das:2021esm} are shown by the unshaded magenta dot-dashed, dashed and dotted lines for $\sqrt{s}= 250$ GeV, 500 GeV and 1 TeV, respectively considering $M_{Z^\prime} >> \sqrt{s}$. 
For a detailed discussion of the methodology to derive these limits, see Ref.~\cite{Das:2021esm}. Note that the limit on $g_1'-M_{Z'}$ varies depending on the values of $x_H$ as couplings between SM particles and $Z'$ also varies. For example, with $x_H =-2$, there is no interaction of $Z'$ with left-handed quarks or leptons, for $x_H =-1$ there is no interaction between right-handed charged-leptons and $Z'$. Similarly, the right-haded down-type quarks have no interaction with $Z'$ for $x_H = 1$. For $x_H=0$ and 2, all the SM particles has non-trivial coupling with $Z'$.
From Fig.~\ref{fig:collider-constraints}, we see that the most stringent constraint up to $M_{Z'}= 6$ TeV comes from LHC dilepton channels. Above $M_{Z'}= 6$ TeV, the resonant $Z'$ production is kinematically limited at $\sqrt{s} = 13$ TeV LHC. Due to this same reason, one does not expect further improvement at HL-LHC. On the other hand, from the projected sensitivities we see that the lepton colliders does better for heavy $Z'$ bosons compare to LHC limit. For the rest of this paper we will consider a specific benchmark value of $M_{Z'}$ and $g_1'$, which is allowed by the current limit.
\subsection{Bounds on the mixing parameter between physical mass eigenstates}
\label{sec:scalar-mix}
\begin{figure}[h!]
\includegraphics[width=0.70\textwidth]{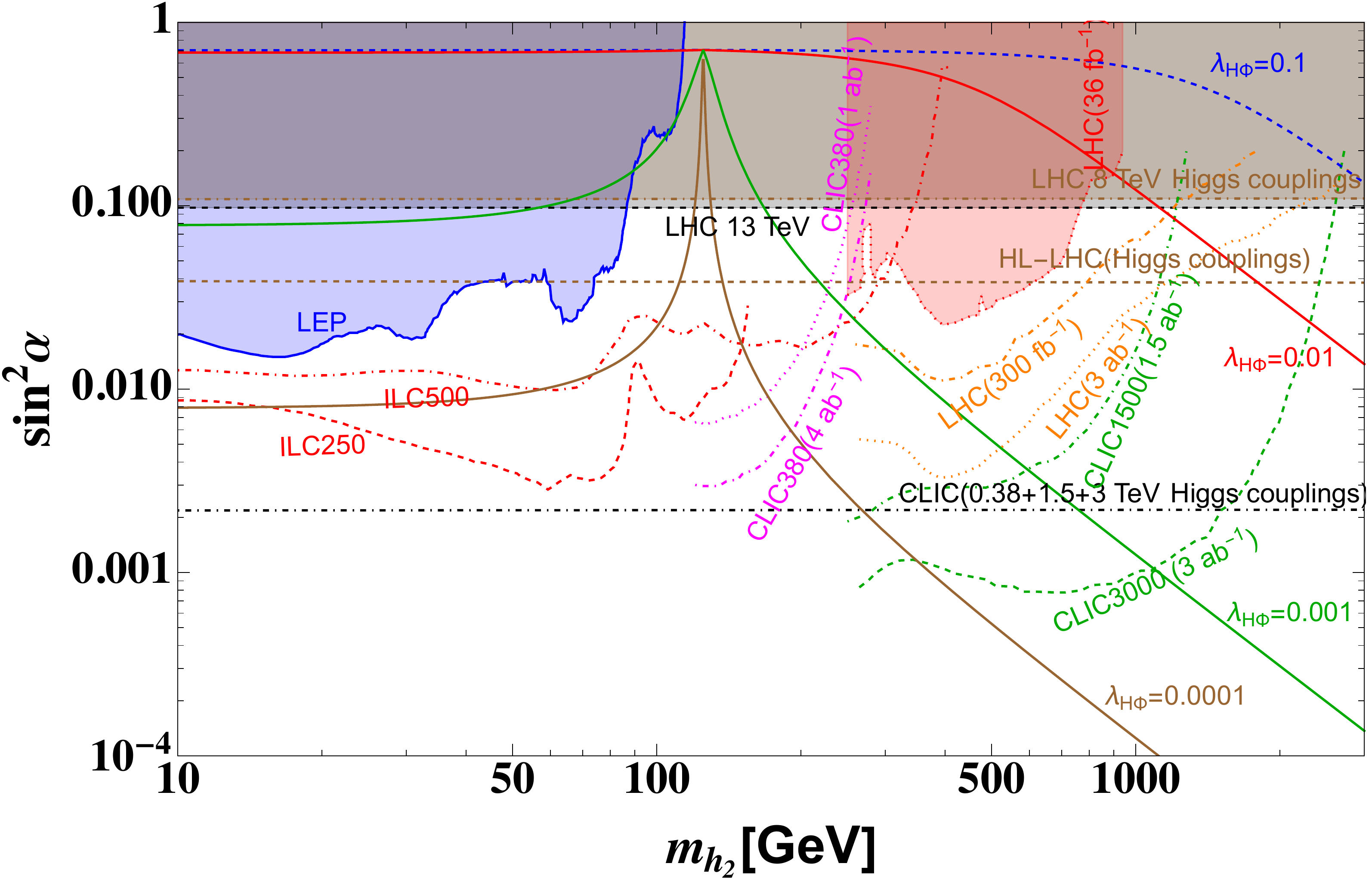}
\caption{Bounds on the mixing parameter $\sin^2\alpha$ between the as a function of $m_{h_2} $from LHC \cite{deBlas:2018mhx}, LEP \cite{LEPWorkingGroupforHiggsbosonsearches:2003ing}, prospective colliders like ILC \cite{Wang:2020lkq} and CLIC \cite{deBlas:2018mhx} respectively. Shaded regions are ruled out by corresponding experiments. Limits on the scalar mixing parameter  from Eq.~\ref{coup} are also shown for different $\lambda_{H\Phi}=$0.1, 0.01, 0.001 and 0.0001 respectively.}
\label{fig:mix}
\end{figure}

We summarize the bounds on the scalar mixing angle $\alpha$ from the LHC \cite{deBlas:2018mhx}, LEP \cite{LEPWorkingGroupforHiggsbosonsearches:2003ing} results, prospective colliders like ILC \cite{Wang:2020lkq} and CLIC \cite{deBlas:2018mhx} in Fig.~\ref{fig:mix}. The prospective limits on the mixing parameter $\sin\alpha$ from the $\sqrt{s}=250$ GeV and 2 ab$^{-1}$ luminosity are shown by the red dashed line with polarization effect $|P(e^+ e^-)|=(30\%, 80\%)$ from the $Z h_2$ mode where $Z$ boson decays to light charged leptons. Corresponding bounds from the $\sqrt{s}=500$ GeV are shown by the red dot-dashed line for $Z\to \mu^+\mu^-$ events. Where as the LEP bounds shown by the blue solid line consider both $Z\to \mu^+\mu^-$ and $e^+e^-$ modes. For direct comparison with the LEP bounds ILC projections can be scaled by a factor of $\frac{1}{\sqrt{2}}$ assuming $Z \to e^-e^+$ channel is similar to $Z \to \mu^- \mu^+$ channel as analyzed in \cite{LEPWorkingGroupforHiggsbosonsearches:2003ing}. Bounds on the scalar mixing parameter from the Higgs couplings at the 8 TeV LHC and High Luminosity LHC (HL-LHC) are shown by the brown dot-dashed and dashed lines respectively. The bounds are taken from \cite{deBlas:2018mhx}. The bounds at the 13 TeV LHC are shown by the black dashed line using the signal strength at $95\%$ C. L. from the ATLAS results \cite{ATLAS:2020qdt}. The corresponding CMS signal strength can be found in \cite{CMS:2020gsy} which gives comparatively weaker bounds. The prospective bounds from the combined CLIC analyses art 380 GeV, 1.5 TeV and 3 TeV are shown by the black dot dashed line. Considering the VEV of the $U(1)_X$ theory at 50 TeV we show the limits on the scalar mixing from the Eq.~\ref{coup} considering four different choices of $\lambda_{H\Phi}$ as $0.1$, $0.01$, $0.001$ and $0.0001$ respectively for $10$ GeV $\leq m_{h_2} \leq 3000$ GeV. 
Bounds on the mixing parameter at $95\%$ C. L. from the $h_2 \to Z Z$ mode at 36 fb$^{-1}$ luminosity is shown by the red dotted line \cite{Buttazzo:2018qqp}. The shaded regions by different colors are ruled out by different experiments. The prospective limits on the mixing parameter form the CLIC at $\sqrt{s}=380$ GeV are shown by the dotted (dot dashed) magenta line for 1 (4) ab$^{-1}$ luminosity from \cite{Mekala:2021uvg}. The prospective limits on the mixing parameter from $h_2 \to hh (4b)$ from CLIC at $\sqrt{s}=$1.5 (3) TeV is shown by green dot dashed (dashed) line from \cite{Buttazzo:2018qqp,deBlas:2018mhx}. The prospective bounds on the mixing parameter estimated form $h_2 \to ZZ$ mode at the LHC at 300 (3000) fb$^{-1}$ luminosity using orange dot dashed (dotted) line from \cite{Buttazzo:2018qqp}.
\section{Phenomenology of dark matter}
\label{sec:dark-matter}
In this section we collect the results of our analysis of DM phenomenology. In our model, two-component DM scenario is possible due to the unbroken $\mathbb{Z}_2\otimes\mathbb{Z}'_2$ symmetry. We choose the $\mathbb{Z}_2$-odd fermion $N_3$ in set one and lightest of the $\mathbb{Z}'_2$-odd scalar $\chi_R$ or $\chi_I$ in set two as DM candidates. In our analysis, we assume scalar DM candidate as $\chi_R$, with the condition $\lambda_\Phi\chi\chi<0$~(the opposite scenario with $\lambda_{\Phi\chi\chi}>0$ would have $\chi_I$ as the DM particle). These two DM candidates must satisfy following two experimental constraints:\\
\begin{itemize}
\item The relic density coming from Planck satellite data~\cite{Planck:2018vyg}
\begin{align}
0.1126\leq \Omega_{\text{DM}} h^2\leq 0.1246.
\label{eq:relic-density}
\end{align}
The total relic abundance of DM in our model is given by the sum of the scalar~($\chi$) and fermion~($N_3$) relic abundances:
\begin{align}
\Omega_{\text{DM}} h^2=\Omega_{\chi}h^2+\Omega_{N_3}h^2
\end{align}
Only for solutions falling exactly within the band given in Eq.~\eqref{eq:relic-density} the totality of the DM can be explained by $\chi$ and $N_3$.
\item Direct detection cross section of DM scattering of nucleon set by various experiments such as XENON1T~\cite{XENON:2018voc}, LUX~\cite{LUX:2016ggv} and PandaX-II~\cite{PandaX-II:2016vec}.
\end{itemize}

In order to calculate all the vertices, mass matrices, tadpole equations etc the model is implemented in the SARAH package~\cite{Staub:2015kfa}. On the other hand the thermal component of the DM relic abundance and the DM-nucleon scattering cross section are determined using micrOMEGAS-5.0.8~\cite{Belanger:2020gnr}. Even though the model introduces new free parameters, not all of them are important to DM analysis. For example, self-quartic coupling $\lambda_{\chi}$ does not play any role in DM phenomenology. Hence we choose to fix $\lambda_\chi=0.1$ in our analysis. The remaining free parameters relevant for DM analysis can be chosen as:
\begin{align}
m_{h_2},\,\, \sin\alpha,\,\, g_1',\,\, M_{Z'},\,\, x_H,\,\, \lambda_{H\chi},\,\, \lambda_{\phi\chi} \text{ and } \lambda_{\phi\chi\chi}.
\end{align} 

In the next sections, we will study how the DM phenomenology of this model depends on the above mentioned parameters and to do that we choose the following benchmark points which are allowed from all the above mentioned constraints:
\begin{align}
\textbf{BP: }m_{h_2}=1~\text{TeV}, \sin\alpha=0.01, g_1'=0.1, M_{Z'}=5~\text{TeV and } x_H=-1.
\label{eq:BP}
\end{align} 
For simplicity, we further assume $\lambda_{H\chi}=\lambda_{\Phi\chi}\equiv\lambda$. We choose $\lambda_{\Phi\chi\chi}$ very small and negative throughout our DM analysis.
\subsection{Relic density}
\begin{figure}[h!]
\includegraphics[width=0.7\textwidth]{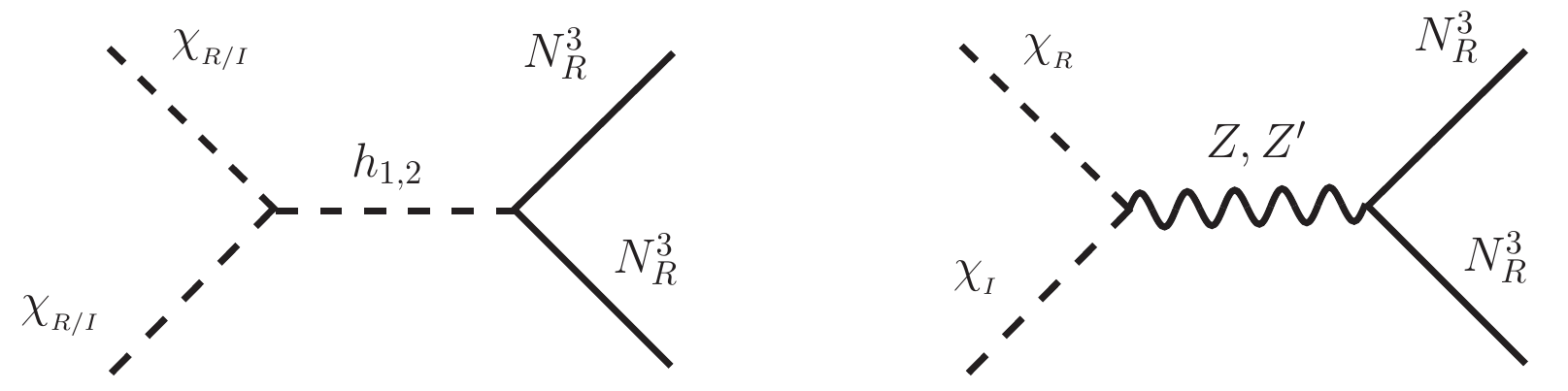}\\
\includegraphics[width=0.7\textwidth]{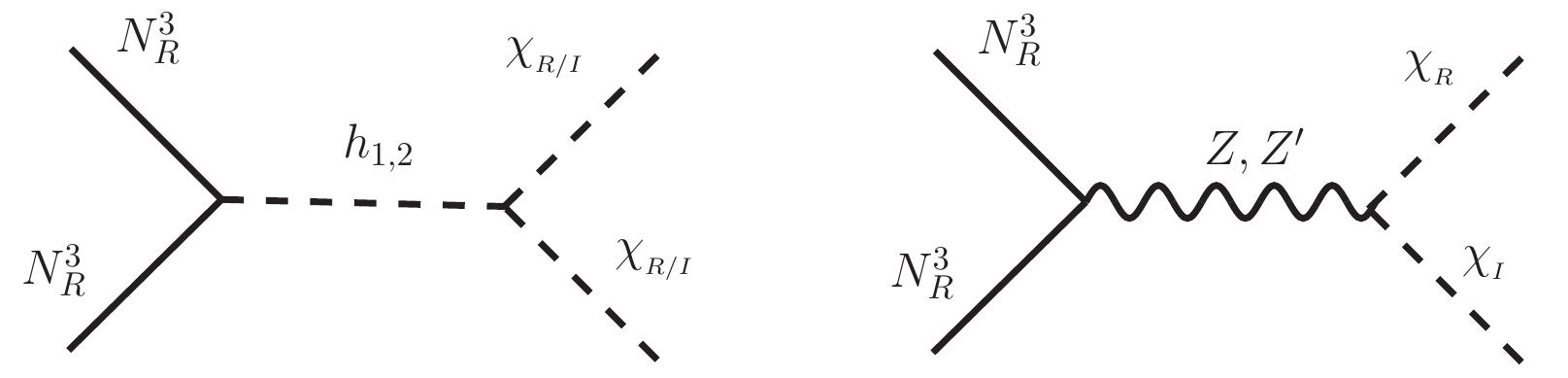}
\caption{The conversion channels which contributes to the two component DM scenario.}
\label{fig:diagram-conversion-annihilation}
\end{figure}

There are several annihilation and co-annihilation diagrams which will contribute to the relic abundances of DM candidates, $\chi$ and $N_3$. We collect all the Feynman diagrams contributing to $\chi_R,\,N_3$ annihilations and co-annihilations in Figs.~\ref{fig:annihilation-diagram-scalar}, \ref{fig:annihilation-diagram-fermion} and \ref{fig:annihilation-diagram-conversion} of Appendix~\ref{app:feynman diagram}. Also in Table.~\ref{tab:cubic-coupling} and \ref{tab:quartic-coupling} of Appendix~\ref{app:vertices}, we have listed the cubic and quartic scalar interactions which play a role in the annihilation channels. We find that the relic density of scalar DM $\chi_R$ is mostly determined by CP-even scalars~($h_{1,2}$) and gauge-bosons~($Z,Z'$)-mediated s-channel annihilation and co-annihilation to SM final states~($\ell^+\ell^-$, $q\bar{q}$, $W^+W^-$, $ZZ$, $h_1h_1$) as well as to $N_{1,2}N_{1,2}$, $h_2h_2$ and $Z'Z'$ final states. A sub-dominant role is played by annihilation into $h_1 h_1,h_2h_2$ and $ZZ,Z'Z'$ via the direct 4-point vertices $h_1^2\chi_{R/I}^2$, $h_2^2\chi_{R/I}^2$ and $Z^2\chi_{R/I}^2$, $Z'^{2}\chi_{R/I}^2$, respectively. Also there could be additional contribution from $\chi_{R/I}$ exchange in the t-channel. The fermionic DM $N_3$ relic density is determined by the $h_{1,2}/Z/Z'$-mediated s-channel annihilation to SM final states as well as to $N_{1,2}N_{1,2}$, $Z'Z'$, $Zh_{1,2}$ and $Z'h_{1,2}$. Besides these above DM annihilation channels, one also needs to take into account the possible conversion of one DM particle into the other, $\chi\chi\leftrightarrow N_3 N_3$. These are shown in Fig.~\ref{fig:diagram-conversion-annihilation}, which are mediated by s-channel $h_{1,2}/Z/Z'$.\\

\begin{figure}[h!]
\includegraphics[width=0.45\textwidth]{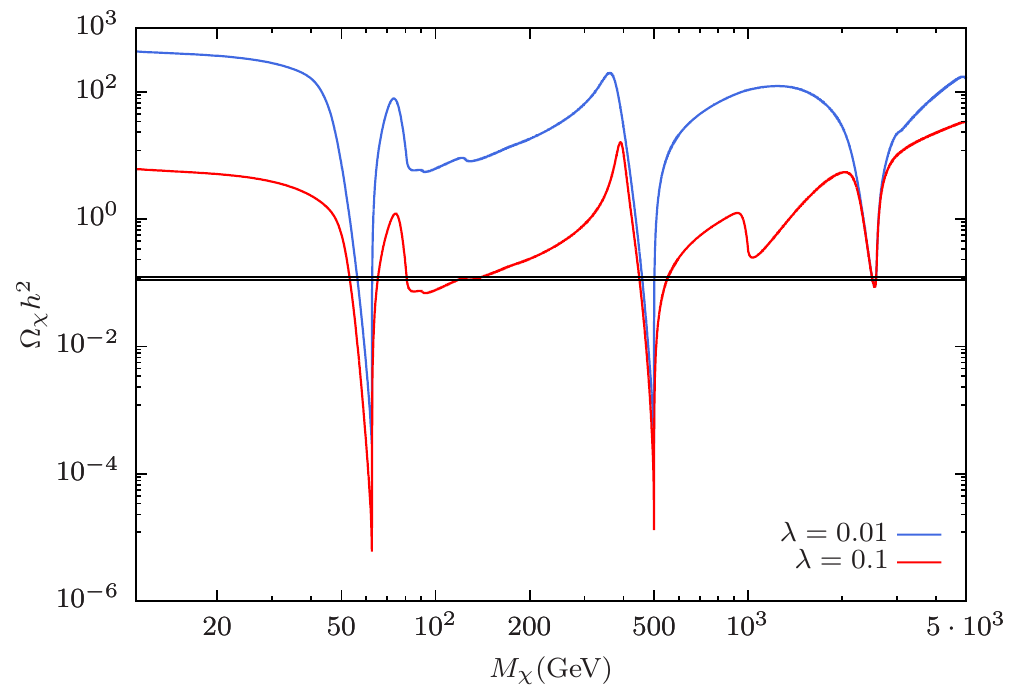}
\includegraphics[width=0.45\textwidth]{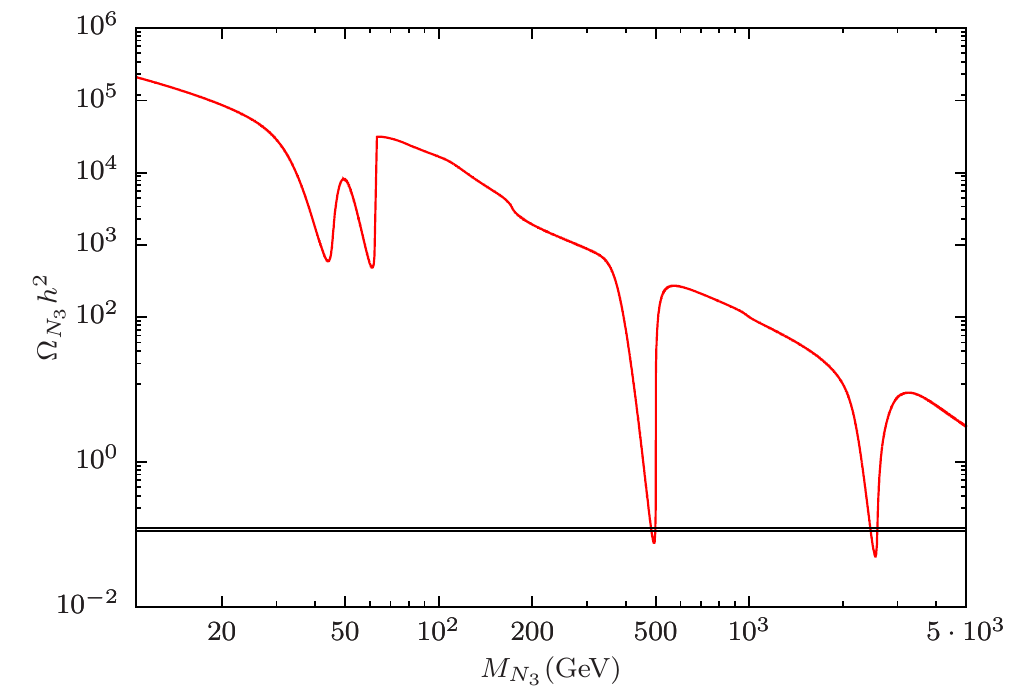}
\caption{Relic density of scalar (right) and fermion~(right) DM as a function of DM mass for one DM candidate. Here, we set the parameter space as \textbf{BP} given in Eq.~\eqref{eq:BP}. Blue and red lines stand for $\lambda=0.01$ and 0.1.}
\label{fig:relic-nocon}
\end{figure}

Firstly, we show the relic density of scalar and fermionic DM in the left and right panel of Fig.~\ref{fig:relic-nocon}, where we neglect the conversion $\chi\chi\leftrightarrow N_3N_3$. In the left panel of Fig.~\ref{fig:relic-nocon} we show the relic density of scalar DM $\chi_R$ for two benchmark points $\lambda=0.01$~(blue line) and 0.1~(red line). We see that there are few dips and the reasons for these dips can be understood by looking in detail into the $\chi_R$ annihilation channels. The dips at $M_{\chi_R}\sim m_{h_1}/2$ and $M_{\chi_R}\sim m_{h_2}/2$ occurs due to annihilation via s-channel $h_1$ and $h_2$ exchange, respectively.
The annihilation becomes very efficient when the Higgs bosons $h_1$ and $h_2$ are on-shell. For $M_{\chi_R}> 90$ GeV, annihilation are dominated by gauge boson final states $W^+W^-$ and $ZZ$, thus explaining the drop at $M_{\chi_R}\sim 90$~GeV. In the mass range $M_{\chi_R}\geq 125$~GeV, $\chi_R$ also annihilate also into SM-like Higgs bosons $h_1 h_1$. Also, when $M_{\chi_R}>m_t$, a new channel $\chi_R\chi_R\to t\bar{t}$ opens up. For DM mass $M_{\chi_R}\geq m_{h_2}$, $\chi_R\chi_R\to h_2 h_2$ channel opens up and this becomes dominant for large $\lambda$, hence the drop in relic density at $M_{\chi_R}\sim m_{h_2}$ for $\lambda=0.1$~(red line). When the DM mass $M_{\chi_R}\sim M_{Z'}/2$, annihilation through the $Z'$-portal becomes efficient, hence the drop at $M_{Z'}/2$. For very heavy DM mass $M_{\chi_R}$, annihilation cross section drops as $\sim 1/M_{\chi_R}^2$, hence the relic density increases. One more important point to note is that as $\chi_R$ and $\chi_I$ mass difference is small due to small value of $\lambda_{\Phi\chi\chi}$, co-annihilation channels with $\chi_I$ occur in all regions of the parameter space, with the effect of lowering the relic density. In the right panel of Fig.~\ref{fig:relic-nocon}, we show the relic density for fermionic DM $N_3$. Again the drops at $m_{h_1,h_2}/2$ and $M_{Z'}/2$ due to s-channel annihilation through on-shell $h_{1,2}$ and $Z'$. Note the presence of annihilation dip at $M_{Z}/2$ for fermionic DM unlike the scalar DM case. The reason behind no dip at $M_Z/2$ for scalar DM is that the $Z$-mediated dip is momentum suppressed. Note that without the conversion $\chi\chi\to N_3 N_3$, annihilation cross section for $N_3$ has no dependence on $\lambda$, hence only one line instead of two lines in the right panel of Fig.~\ref{fig:relic-nocon}.\\
Secondly, we take into account the conversion of two component DM $\chi\chi\leftrightarrow N_3 N_3$, which can be mediated by s-channel $h_{1,2}/Z/Z'$. In this case, we need to simultaneously follow their abundances in the early Universe. The coupled Boltzmann equations are given by~\cite{Esch:2014jpa,Belanger:2012vp}
\begin{align}
& \frac{dY_{\chi}}{dx} = -\sqrt{\frac{45}{\pi}}g^{1/2}_*M_{Pl}\frac{m}{x^2}\Bigg[ \langle \sigma v\rangle^{\chi\chi \rightarrow FF} \left(Y^2_{\chi}-\overline{Y}^2_{\chi}\right) + \langle \sigma v\rangle^{\chi\chi\rightarrow N_3 N_3} \left(Y^2_{\chi}-\overline{Y}^2_{\chi}\frac{Y^2_{N_3}}{\overline{Y}^2_{N_3}}\right) \Bigg] \\
& \frac{dY_{N_3}}{dx} = -\sqrt{\frac{45}{\pi}}g^{1/2}_*M_{Pl}\frac{m}{x^2}\Bigg[ \langle \sigma v\rangle^{N_3 N_3 \rightarrow FF} \left(Y^2_{N_3}-\overline{Y}^2_{N_3}\right) + \langle \sigma v\rangle^{N_3 N_3\rightarrow \chi\chi} \left(Y^2_{N_3}-\overline{Y}^2_{N_3}\frac{Y^2_{\chi}}{\overline{Y}^2_{\chi}}\right) \Bigg]
\end{align}
here $x=\frac{m}{T}$ and $m=\frac{M_{\chi}+M_{N_3}}{2}$. $\langle \sigma v\rangle$ is the thermally average annihilation cross-section.
\begin{figure}[h!]
\includegraphics[width=0.45\textwidth]{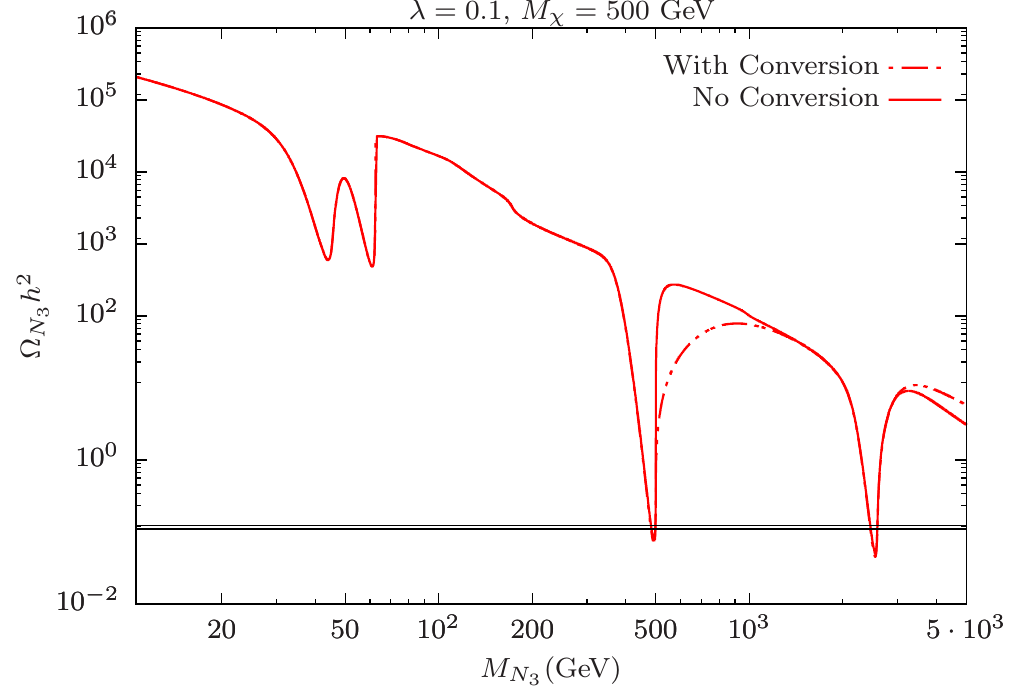}
\includegraphics[width=0.45\textwidth]{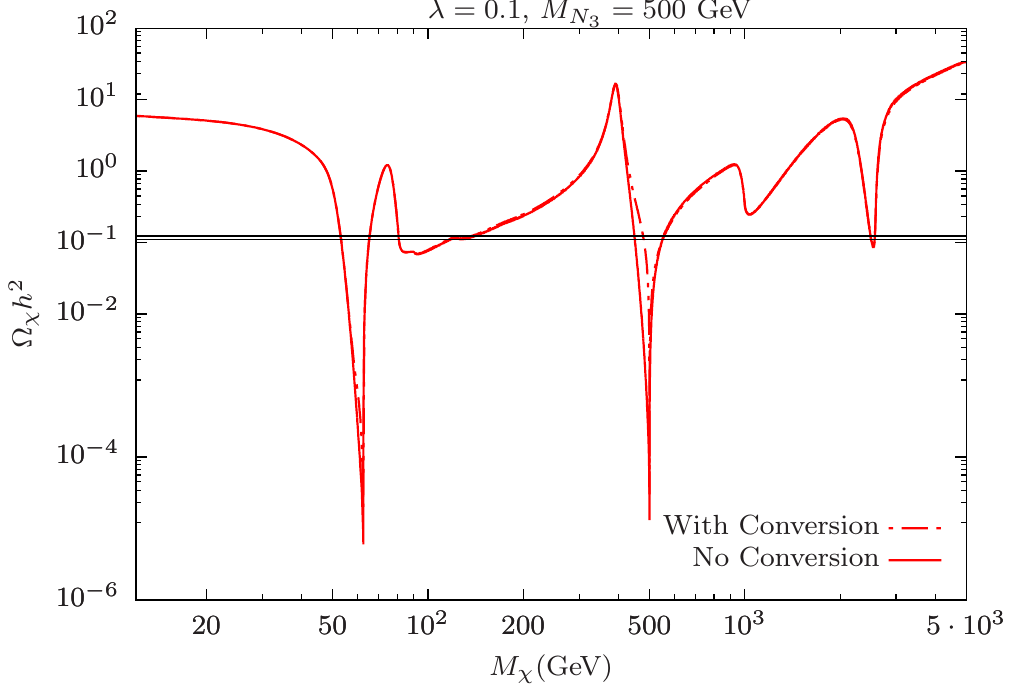}
\caption{Effect of conversion channel $\chi\chi\leftrightarrow N_3 N_3$ on the relic density. The solid and dashed lines stand for the case of without and with the conversion channel. Left(right) panel is for fermionic(scalar) DM with $M_{\chi}=500$~GeV~($M_{N_3}=500$~GeV) and $\lambda=0.1$. The other parameters are fixed as \textbf{BP} given in Eq.~\eqref{eq:BP}.}
\label{fig:relic-con}
\end{figure}
In the above $F$ can be any particle except $\chi$ and $N_3$. In these equations $g^{1/2}_*$, $M_{Pl}$ and $\overline{Y}$ are degrees of freedom, Planck mass and equilibrium value of $Y$, respectively. In above equations, the most right term is responsible for the conversion process $\chi\chi \leftrightarrow N_3 N_3$. As $\chi\chi\to N_3 N_3$ and $N_3 N_3\to\chi\chi$ are determined by the same squared matrix elements~(see Fig.~\ref{fig:diagram-conversion-annihilation}), they are not independent but related to each other. These conversion processes are mediated by $h_{1,2}/Z/Z'$. Note that the couplings $\chi_R-\chi_I-Z$ and $N_3-N_3-Z$ are suppressed by $\sin\theta'$. Hence, the conversion process mediated by $h_{1,2}/Z'$ gives the dominant contribution, provided $\lambda_{H\chi}$, $\lambda_{\Phi\chi}$, $\sin\alpha$ and $g_1'$ are not too small. The effect of conversion on the $N_3/\chi$ relic density is shown in the left and right panel of Fig.~\ref{fig:relic-con}. Solid and dashed line represents the relic for without-conversion and with-conversion cases. In both panels, we fixed the other DM mass as $M_{\chi}(M_{N_3})=500$~GeV, where as fixing other parameters same as before. For the fermionic DM, when $M_{N_3}>M_{\chi}$, the larger the quartic coupling $\lambda$, the larger the annihilation rate $N_3 N_3\to \chi\chi$ and hence smaller the relic density. Note that, when $M_{N_3}< M_{\chi}$, the effect of conversion $N_3 N_3\to\chi\chi$ is very small. Also as for conversion case, $Z'\to \chi\chi$ process contributes to the total decay width of $Z'$, it can greatly enhance the total decay width of $Z'$, which causes the increase of $\Omega h^2$ above $M_{N_3}\sim M_{Z'}/2$ for small mass of $M_{\chi}$. In addition to this, for $M_{\chi}<m_{h_2}/2$, there can be modification in $\Omega h^2$ due to the change in the $h_2$ decay width. For the scalar DM, the annihilation channel $\chi\chi\to N_3 N_3$ via $Z$ and $Z'$ are momentum suppressed. The dominating contribution to the conversion comes from $\chi\chi\to N_3 N_3$ annihilation through $h_{1,2}$ mediated processes. But as $h_1-N_3-N_3$ coupling is suppressed by $\sin\alpha$, we see that only changes occurs around $\sim m_{h_2}/2$ for $M_{N_3}=500$~GeV.\\
\begin{figure}[h!]
\includegraphics[width=0.45\textwidth]{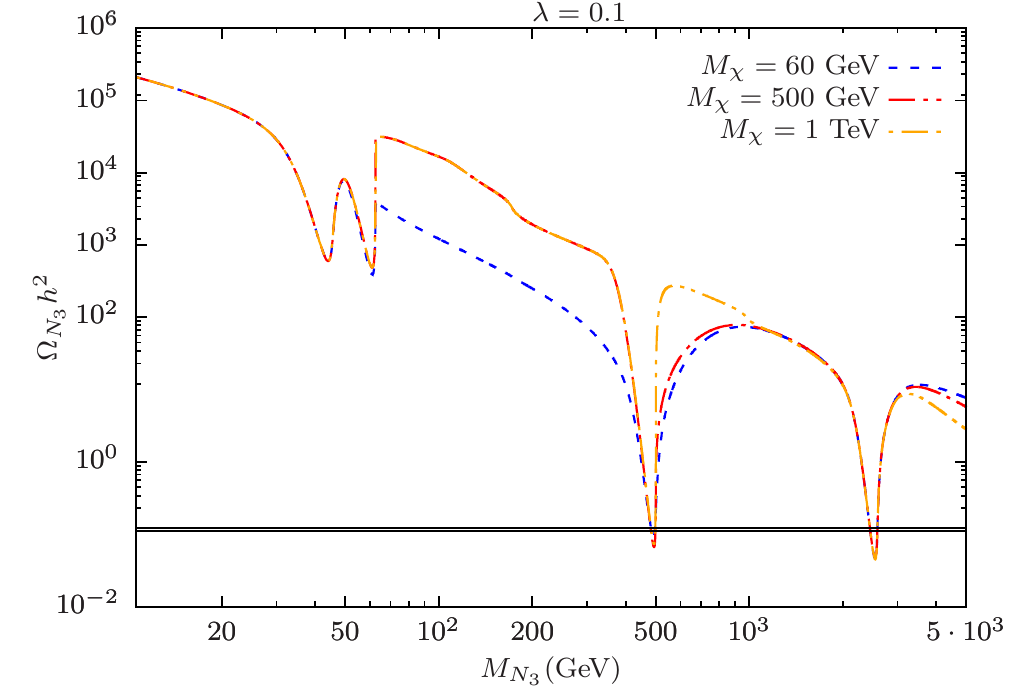}
\includegraphics[width=0.45\textwidth]{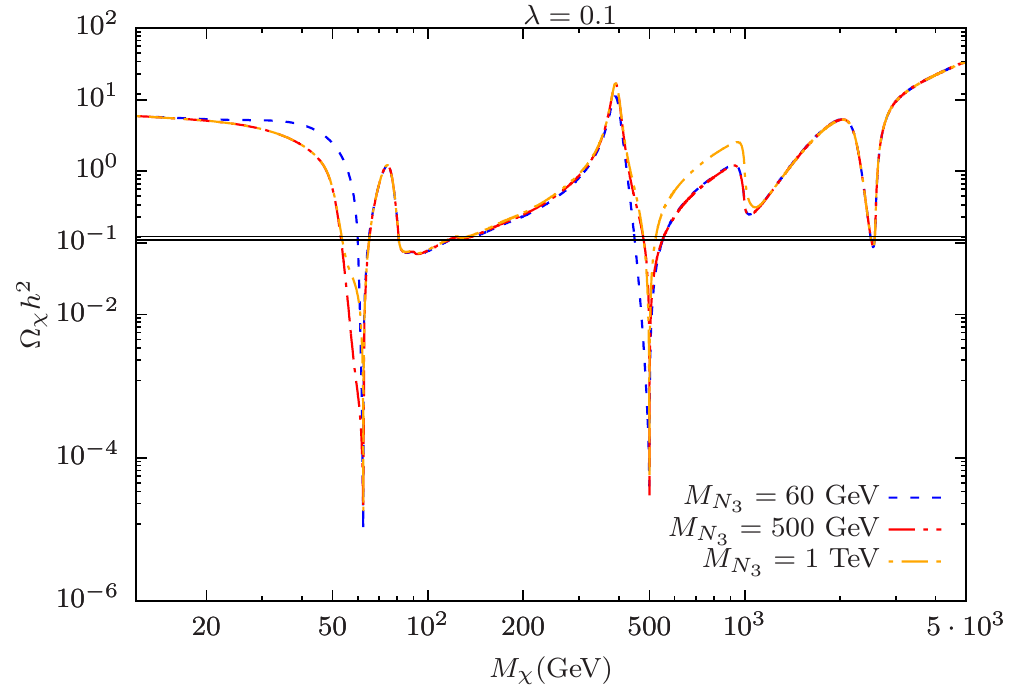}
\caption{The effect of two component DM conversion on $N_3$~(left panel) and $\chi$~(right panel) relic density for fixed $\lambda = 0.1$. The other parameters are fixed as \textbf{BP} given in Eq.~\eqref{eq:BP}.}
\label{fig:relic-mass}
\end{figure}
In Fig.~\ref{fig:relic-mass}, we extended the study of conversion effect and two DM boltzmann equations. The left(right) of Fig.~\ref{fig:relic-mass} shows the $N_3(\chi)$ relic density for $M_{\chi(N_3)}=60, 500$~GeV and 1 TeV with fixed value of $\lambda=0.1$. In case of fermionic DM, it is clear that for small $M_{\chi}$ the annihilation cross section $N_3 N_3\to \chi\chi$ is large and hence the smaller relic density for $M_{\chi}=60$~GeV compare to other masses. For relatively heavy scalar DM mass, such as $M_{\chi}=500$~GeV and 1 TeV, the conversion effects on the $N_3$ relic density are almost negligible when $M_{N_3}<m_{h_2}/2$. Since, the coupling $h_2-N_3-N_3$ is directly proportional to $M_{N_3}$, in the high mass region, the conversion effect is comparable to $h_2-N_3-N_3$ coupling effect. This makes the dependence of relic density on $M_{\chi}$ nonlinear in the high $M_{N_3}$ region. On the other hand, for scalar DM the dependence of relic density on fermionic DM mass is simple. As we previously mentioned, the conversion effects mainly comes from the channel $\chi\chi\to N_3 N_3$ via $h_1/h_2$. Hence we only see some changes for different $M_{N_3}$ masses either around $M_{\chi}\sim m_{h1}/2$ or around $M_{\chi}\sim m_{h_2}/2$. Also if $M_{N_3}<m_{h_{1,2}}/2$, the $h_{1,2}$ decay width changes, which in turn modifies the relic. In conclusion, $h_i-N_3-N_3$ and $h_i-\chi\chi$ couplings plays important role in two component DM conversion. When $M_{\chi}\sim M_{N_3}$, the conversion can take place in both direction, if not the case, only the conversion of heavier one into lighter one is important.
\subsection{Direct detection}
Let us now study the direct detection prospects of our DM candidates $\chi_R$ and $N_3$. The current experimental constraints on the DM direct detection assume the existence of only one DM candidate. As in our model two-component DM candidates are predicted, the contribution of each candidate to the direct detection cross section should be rescaled by the fraction contributing to the total relic density. Hence it is convenient to define the fraction of the mass density of $i-$th DM in case of multi-component DM~\cite{Cao:2007fy,Wang:2015saa,Aoki:2012ub,Bhattacharya:2013hva}
\begin{align}
\epsilon_i=\frac{\Omega_i h^2}{\Omega_{\text{DM}} h^2}
\label{eq:fraction}
\end{align}
The upper limit on the direct detection now can be recasted as 
\begin{align}
\frac{\epsilon_\chi}{M_\chi}\sigma_{\chi-N}+\frac{\epsilon_{N_3}}{M_{N_3}}\sigma_{N_3-N}<\frac{\sigma^{\text{exp}}}{M_{\text{DM}}}
\label{eq:direct-detection}
\end{align}
where, $\sigma_{\chi-N}$ and $\sigma_{N_3-N}$ are the scattering cross section of $\chi$ and $N_3$ with nucleon $N$.\\

\begin{figure}[h]
\centering
\includegraphics[width=0.25\textwidth]{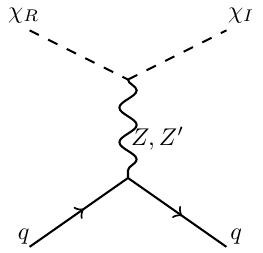}~~~~~~~~
\includegraphics[width=0.25\textwidth]{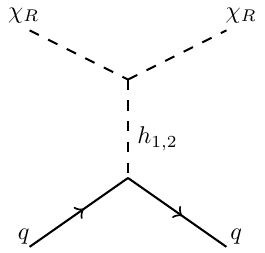}
\caption{\footnotesize{$Z,Z'$ and Higgs-mediated tree-level Feynman diagrams contributing to the scattering of $\chi_R$ off nuclei.} }
\label{fig:scalar-dd-feynman}
\end{figure}

\begin{figure}[h]
\centering
\includegraphics[width=0.25\textwidth]{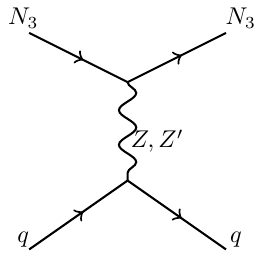}~~~~~~~~
\includegraphics[width=0.25\textwidth]{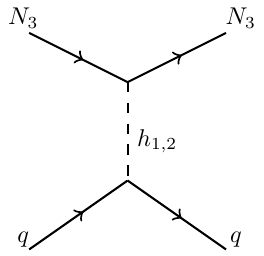}
\caption{\footnotesize{$Z,Z'$ and Higgs-mediated tree-level Feynman diagrams contributing to the scattering of $N_3$ off nuclei.} }
\label{fig:fermionic-dd-feynman}
\end{figure}

\begin{figure}[h!]
\includegraphics[width=0.45\textwidth]{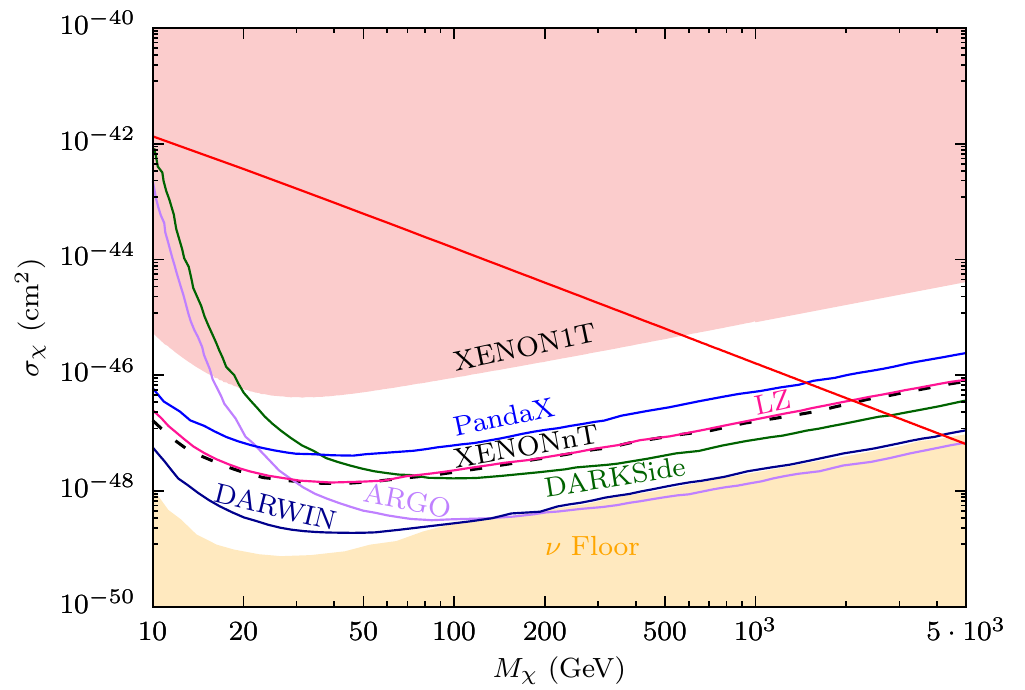}
\includegraphics[width=0.45\textwidth]{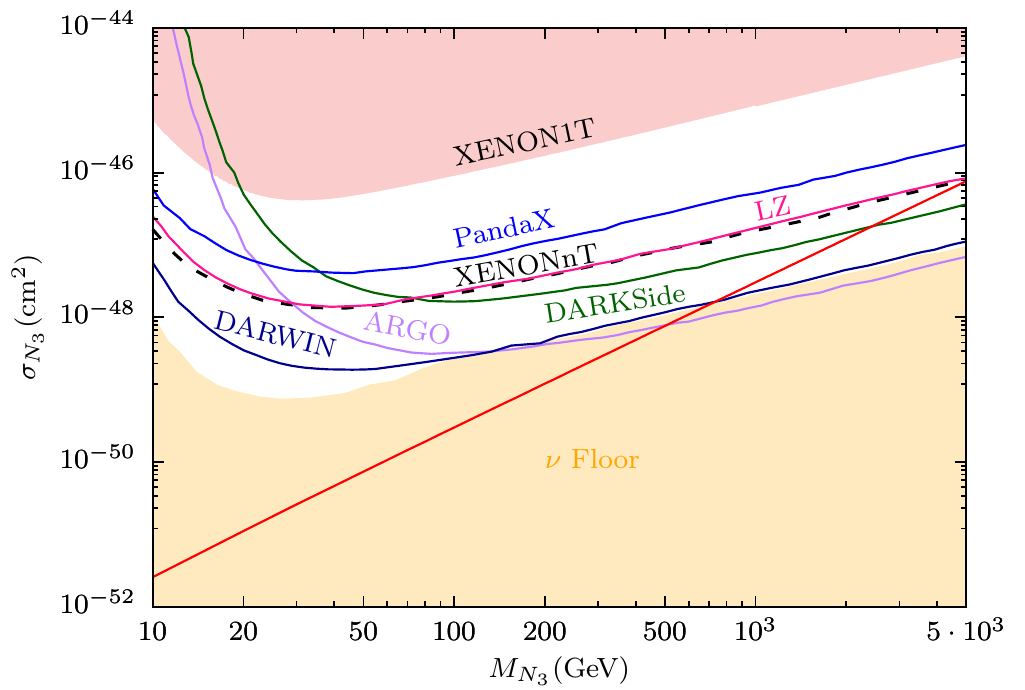}
\caption{Spin-independent DM-nucleon elastic scattering cross section versus the DM mass. Left and right panel stands for scalar and fermionic DM case. The light-red shaded region denotes the excluded region coming from the XENON1T experiment~\cite{Aprile:2018dbl}. The light-orange region corresponds to the “neutrino floor” coming from coherent elastic neutrino scattering~\cite{Billard:2013qya}. We have also shown various projected sensitivities coming from experiments such as PandaX-4t~\cite{Zhang:2018xdp}, LUX-ZEPLIN(LZ)~\cite{Akerib:2018lyp}, XENONnT~\cite{Aprile:2020vtw}, DarkSide-20k~\cite{DS_ESPP}, DARWIN~\cite{Aalbers:2016jon} and ARGO~\cite{Billard:2021uyg}.}
\label{fig:dd}
\end{figure}

\underline{\bf Estimation of $\sigma_{\chi-N}$: }
In this model, the scattering of the scalar DM candidate $\chi_R$ with a nucleon happens via two t-channel diagrams with either $Z,Z'$ or $h_{1,2}$ as propagators, shown in Fig.~\ref{fig:scalar-dd-feynman}. Notice that, as the complex scalar $\chi$ has non-zero $U(1)_X$ charge, the $\chi_R$-nucleon spin-independent~(SI) cross-section can be mediated by the $Z,Z'$-boson. Generally this exceeds the current limit from direct detection experiments like XENON1T. However this can be easily avoided by taking non-zero $\lambda_{\Phi\chi\chi}$. In this case there is a small mass splitting between $\chi_R$ and $\chi_I$, so that the interaction through the $Z,Z'$-boson is kinematically forbidden or leads to inelastic scattering. As a result, for nonzero $\lambda_{\Phi\chi\chi}$, the $\chi_R$-nucleon interaction via the Higgs~($h_{1,2}$) will be the dominant one. The effective lagrangian for nucleon-DM interaction can be written as
\begin{align}
\mathcal{L}_{\text{eff}}=a_N\bar{N}N\chi_R^2
\end{align}
where $a_N$ is the effective coupling between DM and nucleon. The resulting spin independent scattering cross section is given by
\begin{align}
\sigma^{\text{SI}}_{\chi-N}=\frac{\mu_N^2 m_N^2 f_N^2}{4\pi M_{\chi_R}^2 v_H^2}\Big(\frac{\lambda_{h_1\chi_R\chi_R}}{m_{h_1}^2}\cos\alpha - \frac{\lambda_{h_2\chi_R\chi_R}}{m_{h_2}^2}\sin\alpha\Big)^2,
\label{eq:SI-scalar}
\end{align}
where $\mu_N=\frac{m_N M_{\chi_R}}{m_N+M_{\chi_R}}$ is the reduced mass for nucleon-DM system. Here $f_N$ is the form factor, which depends on hadronic matrix elements. The trilinear couplings $\lambda_{h_1\chi_R\chi_R}$ and $\lambda_{h_2\chi_R\chi_R}$ are given as 
\begin{align}
\lambda_{h_1\chi_R\chi_R}&=\lambda_{H\chi}v_H \cos\alpha + \lambda_{\Phi\chi}v_\Phi\sin\alpha + \sqrt{2}\lambda_{\Phi\chi\chi}\sin\alpha \\
\lambda_{h_2\chi_R\chi_R}&=-\lambda_{H\chi}v_H \sin\alpha + \lambda_{\Phi\chi}v_\Phi\cos\alpha + \sqrt{2}\lambda_{\Phi\chi\chi}\cos\alpha 
\end{align}
The above formula in Eq.~\eqref{eq:SI-scalar} is an extension of the expression corresponding to the singlet scalar DM case~\cite{Cline:2013gha}. The relative negative sign between the $h_1$ and $h_2$ contribution arises as in our considered model as the coupling of the Higgs boson to \sm particles gets modified according to the substitution rule given in Eq.~\eqref{eq:substitution}. Due to the presence of the two different channels, depending on the parameter space we can have destructive interference between these two channels and direct detection can be very small.\\
\underline{\bf Estimation of $\sigma_{N_3-N}$: }
Again for the fermionic case also there will be contributions from t-channel diagrams with either $Z,Z'$ or $h_{1,2}$ as propagators, shown in the left and right panel of Fig.~\ref{fig:fermionic-dd-feynman}. The $Z,Z'$-mediated diagram contributions to spin independent cross section are velocity suppressed and hence remain within the experimental bounds~\cite{Arcadi:2020aot}. The Higgs-mediated contribution to spin independent cross section can saturate the current experimental bounds. 
This can be written as~\cite{Lopez-Honorez:2012tov}:
\begin{align}
\sigma^{\text{SI}}_{N_3-N}=\frac{(y_M^3)^2 \mu_N^2 m_N^2 f_N^2}{2\pi v_H^2}\sin^2(2\alpha)\Big(\frac{1}{m_{h_1}^2}-\frac{1}{m_{h_2}^2}\Big)^2
\end{align}
where $\mu_N=\frac{m_N M_{N_3}}{m_N+M_{N_3}}$ is the reduced mass for nucleon-DM system.\\
In Fig.~\ref{fig:dd}, the direct detection limit is shown for each DM separately. In both panels, axes are on logscale which implies a linear behavior as expected from the DM-nucleon cross section for both DM. The DM-nucleon cross section is dominated by higgs mediated channels. The fermionic cross section is smaller due to $\sin^2(2\alpha)$ dependence. The black lines in each panel denotes the latest upper bound from the XENON1T collaboration~\cite{Aprile:2018dbl}. There are constraints from other experiments as well, such as LUX~\cite{Akerib:2016vxi} and PandaX-II~\cite{Tan:2016zwf}, but weaker when compared to the XENON1T limit. There are also the projected sensitivities for the PandaX-4t~\cite{Zhang:2018xdp}, LUX-ZEPLIN(LZ)~\cite{Akerib:2018lyp}, XENONnT~\cite{Aprile:2020vtw}, DarkSide-20k~\cite{DS_ESPP}, DARWIN~\cite{Aalbers:2016jon} and ARGO~\cite{Billard:2021uyg} experiments which we also show in Fig.~\ref{fig:dd}. The lower limit corresponding to the ``neutrino floor'' from coherent elastic neutrino scattering~\cite{Billard:2013qya} is indicated in orange line. We see from Fig.~\ref{fig:dd} that there might be low-mass solutions with the correct DM relic density. However, most of these are ruled out by the XENON1T direct detection cross section upper limits. We see that beyond 500 GeV scalar DM satisfy XENON1T bound, where as fermionic DM satisfy the XENON1T bound for the whole DM mass we have considered, although the low fermionic DM mass is in conflict with Neutrino floor.\\
\begin{table}[ht!]
\setlength\tabcolsep{0.25cm}
\centering
\begin{tabular}{| c | c | c | c | c |}
\hline
$M_{\chi}$ (GeV) & $M_{N_3}$(GeV) & $\epsilon_{\chi}$  &  $ \epsilon_{N_3}$  & $\sum_{\{i=\chi,N_3\}} \epsilon_i \frac{\sigma_{i-N}}{M_i}$ (cm$^2$-GeV$^{-1}$)\\
\hline
500.14 &  499.0  &  0.001268 & 0.99874 &  3.138$\times10^{-51}$\\
62.17 &  2586.11  &  0.000083 & 0.99991 & 6.32$\times10^{-50}$\\
499.65 &  2585.96  &  0.00029 & 0.9997 & 8.27$\times10^{-51}$\\
500.78 &  2585.095  &  0.0043 & 0.9956 & 1.335$\times10^{-50}$\\
61.55 &  2585.348  &  0.00027 & 0.9997 & 1.936$\times10^{-49}$\\
\hline
\end{tabular}
\caption{Benchmark points where both relic abundance and direct detection limit for two component DM are satisfies.}
\label{tab:DD limit}
\end{table}
In table.~\ref{tab:DD limit}, we have shown few benchmark masses which satisfies both the relic abundance and the direct detection limit for our two components DM case. We have listed few data points around which many points can be found that satisfy relic density bound also. In conclusion, we can have parameter space, specially around the resonance regions, where one can satisfy both the relic abundance and direct detection limit. Note that for our choice of benchmark $m_{h_2} = 1$ TeV, the mass fraction is very small, i.e $\epsilon_\chi = \mathcal{O}(10^{-3})$. This is due to the tight constraint coming from direct detection constraints for scalar dark matter, see left panel of Fig.~\ref{fig:dd}. This constraint becomes loose for large mass $M_\chi$, hence for a different benchmark such as $m_{h_2} = 2$ TeV, the mass fraction $\epsilon_\chi$ can be large near the resonance region $M_\chi\sim m_{h_2}/2$ where one can have correct relic density.
\section{Relic density dependence on $U(1)_X$ charge $x_H$}
\label{sec:relic-dependence-xH}
In this model, there are basically two ways for the DM to interact with the SM particles. Either through the Higgs boson interactions or through the $Z'-$boson interactions as all particles in our model are charged under $U(1)_X$. Hence the relic density for both the DM will have some dependence on $U(1)_X$ charge $x_H$ when one considers the $Z'$-portal DM. Fig.~\ref{fig:relic-nocon}-\ref{fig:relic-con} and \ref{fig:relic-mass} also indicates that the $Z'$ boson resonance effect is very important in reproducing the known DM relic abundance and hence, $M_{\text{DM}}\sim M_{Z'}/2$. Hence, in the case of pure $Z'$-portal DM scenario~($\sin\alpha = 0$), the resultant DM relic abundance is controlled by four free parameters, namely, $g_1'$, $M_{Z'}$, $M_{\text{DM}}$ and $x_H$.
\begin{figure}[h!]
\includegraphics[width=0.45\textwidth]{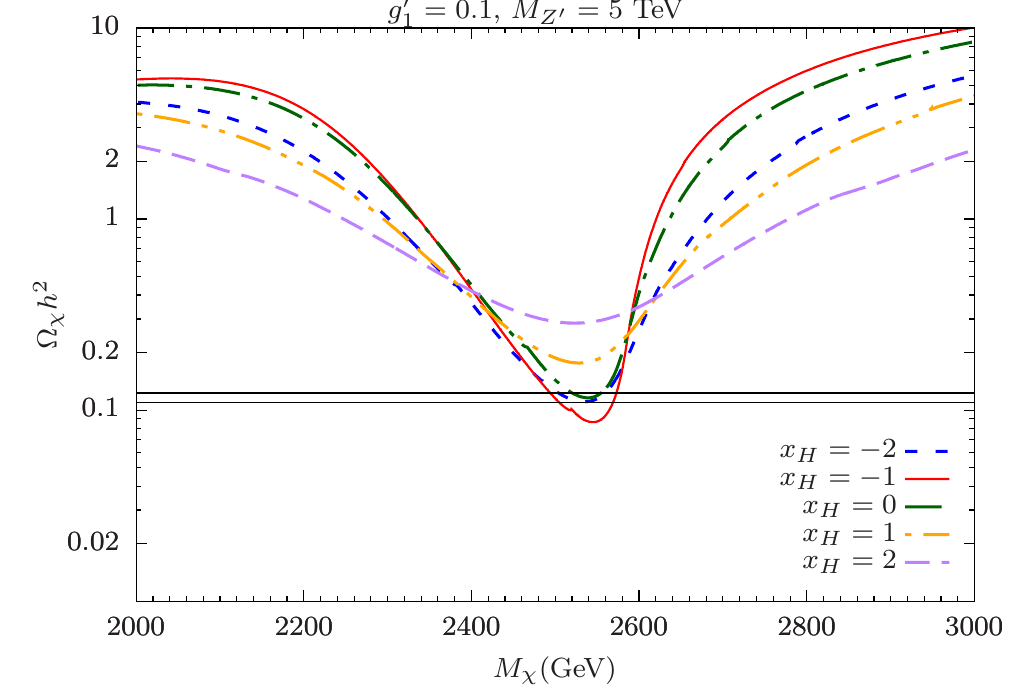}
\includegraphics[width=0.45\textwidth]{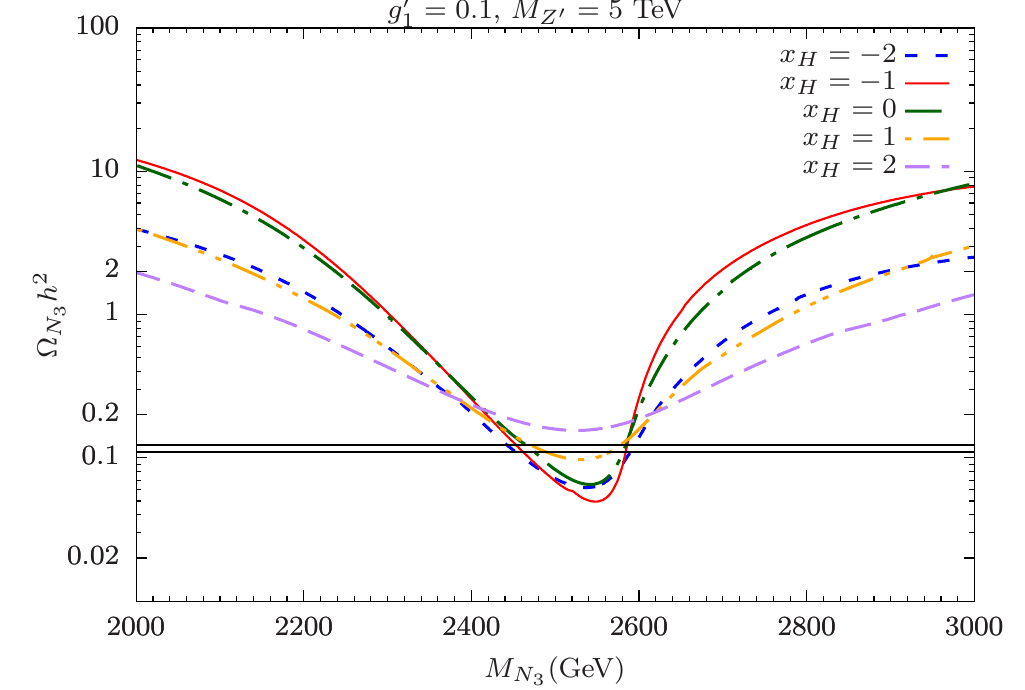}
\caption{The relic density variation is shown with respect to mass of the scalar and fermion DM in left and right panel respectively for different choices of $x_H$. We have fixed $M_{Z'}=5$ TeV and $g_1'=0.1$.}
\label{fig:relic-xH-scalar-fermion}
\end{figure}
In Fig.~\ref{fig:relic-xH-scalar-fermion}, we have chosen multiple $x_H=-2,-1,0,1,2$ etc, to see how relic behave in each of these case for both scalar~(left panel) and fermion~(right panel) DM. We have fixed $M_{Z'}=5$ TeV and $g_1'=0.1$. We have chosen DM masses in the range from 2 to 3 TeV which is most suitable for studying $x_H$ behaviour as we fixed $M_{Z'}=5$ TeV. The dominating channels are here $\chi_R \chi_I/ N_3 N_3 \rightarrow Z'\rightarrow f\bar{f}$, where $f$ is the SM final states. Hence, the DM annihilation cross section for $M_{N_3},M_{\chi}\sim M_{Z'}/2$ is proportional to $1/\Gamma_{Z'}$. On the other hand the $Z'$ decay width depends on the value of $x_H$ as $Z'$ interactions depends on $x_H$ when $x_\Phi$ is fixed. We found that $Z'$ becomes minimum at $x_{H}=-0.8$ and is a symmetric function of $x_H$ around this point. Due to this reason, $-2\leq x_H\leq 0$ are more suitable for both candidate to be DM in $Z'$ portal framework.\\
\begin{figure}[h!]
\includegraphics[width=0.45\textwidth]{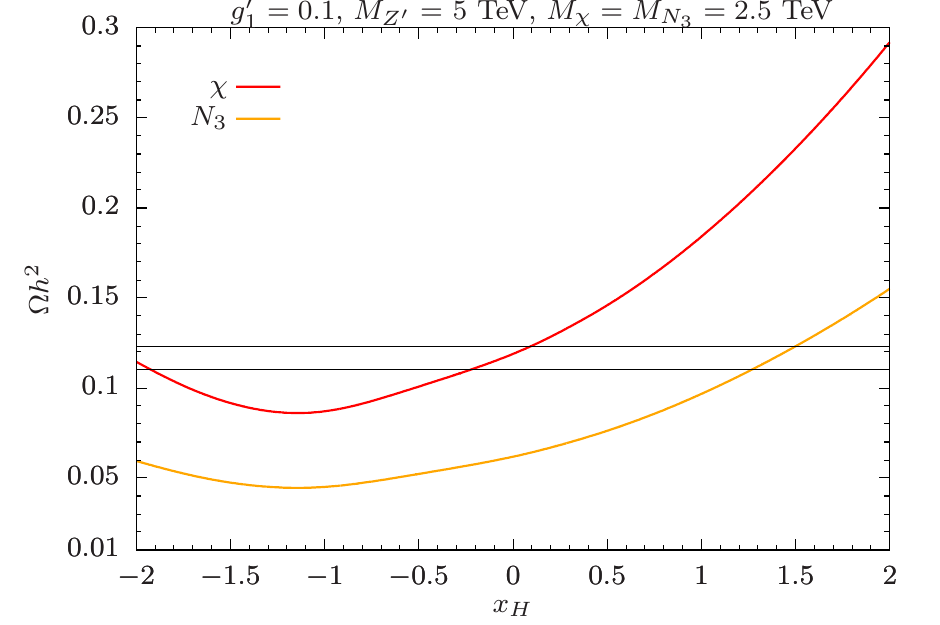}
\includegraphics[width=0.45\textwidth]{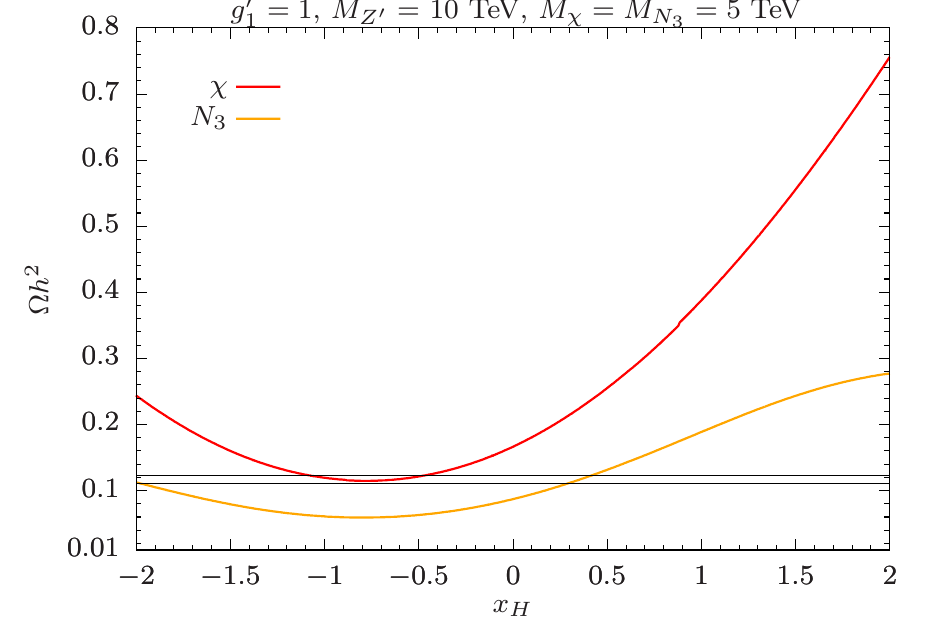}
\caption{The variation of relic density with $U(1)_X$ charge $x_H$ is shown here for $M_{Z'}=5$ TeV~(left panel) and $10$ TeV~(right panel). The red and orange curves are for scalar and fermion DM, respectively. We have set the mass for both the DM to be at $M_{Z'}/2$.}
\label{fig:relic-xH}
\end{figure}
In Fig.~\ref{fig:relic-xH}, we are showing the variation of scalar and fermion DM relic density with respect to $U(1)_X$ charge $x_H$ for $M_{Z'}=5$ TeV~(left panel) and $10$ TeV~(right panel). The mass of both DM candidates have been chosen at $M_{Z'}/2$ since only their relic abundance can have correct order. For heavier $Z'$ mass $M_{Z'}=10$ TeV we choose relatively large coupling $g_1'=1$ which is allowed for any value of $x_H$ from the current collider constraints, see Fig.~\ref{fig:collider-constraints}. From Fig.~\ref{fig:relic-xH}, we again see that $-2\leq x_H\leq 0$ are preferable for both candidates to be DM in $Z'$ portal framework and for larger $M_{Z'}$ one need relatively large coupling $g_1'$ to have correct relic density.
\section{Conclusion}
\label{sec:conclusion}
We have considered a generic $U(1)_X$ model which provides an economical extension of SM to accommodate DM and active neutrino masses. In this generic $U(1)_X$ model, the charges of the SM particles are defined as a linear combination of the SM $U(1)_Y$ and $U(1)_{B-L}$ charges. In addition to three generation of RHNs, we add two complex scalar $\Phi$, $\chi$ both are charged under the $U(1)_X$ gauge group. $U(1)_X$ symmetry breaking is driven by the VEV of $\Phi$. Note that, although Majorana neutrino masses for all three RHNs are generated through the VEV of $\Phi$, but due to the $\mathbb{Z}_2$ charge assignment, only $N_R^{1,2}$ have Dirac Yukawa couplings with the SM lepton doublets. Hence, in this model only two SM neutrinos are massive as the seesaw mechanism generates the SM neutrino mass matrix with only the two Majorana RHNs. As, additional fields such as $N_R^3$ and $\chi$ are odd under the discrete symmetries $\mathbb{Z}_2$ and $\mathbb{Z}_2'$, respectively, they both are stabilized and hence play the role of two component DM. We have discussed the main features of this two-component DM scenario in the context of generic $U(1)_X$ model.

Apart from the SM Higgs boson, we have one more heavy Higgs and an additional neutral gauge boson $Z'$. As a result, there are two ways for both the scalar and fermion DM to interact with the SM particles. One is through the $Z'$ boson interaction since all particles in this model are charged under the gauge group $U(1)_X$. The other is through the Higgs boson interactions. In the case of Higgs-portal DM scenario, we find that SM Higgs and heavy Higgs mixing $\sin\alpha$, heavy Higgs mass $m_{h_2}$ and quartic couplings $\lambda$ plays a crucial role. On the other hand for $Z'$-portal DM scenario, the DM phenomenology is basically controlled by very few parameters such as $g_1'$, $M_{Z'}$, $M_{\text{DM}}$ and $U(1)_X$ charge $x_H$. We also find that, in order to reproduce the observed DM relic density it is required to fix the DM mass around $M_{Z'}/2$ for the case of $Z'$-portal DM. We further found that with the current collider constraint on $g_1'$ and $M_{Z'}$, $-2\leq x_H\leq 0$ are preferable to have the correct relic density.
\begin{acknowledgements}
The work of S.M. is supported by KIAS Individual Grants (PG086001) at Korea Institute for Advanced Study.
\end{acknowledgements}
\appendix
\section{Anomaly cancellations}
\label{anomaly}
In this setup we consider a minimal $U(1)_X$ extension of the SM with a general charge assignment. 
We deduce the gauge and mixed gauge-gravity anomaly cancelation conditions on the $U(1)_X$ charges as follows:
The charge assignments for the fermions are independent of the generation in this scenario. 
Hence we use these charges to deduce the gauge and mixed gauge-gravity anomaly cancelation conditions: 
\begin{align*}
{U}(1)_X \otimes \left[ {SU}(3)_c \right]^2&\ :&
			2x_q - x_u - x_d &\ =\  0, \nonumber \\
{U}(1)_X \otimes \left[ {SU}(2)_L \right]^2&\ :&
			3x_q + x_\ell &\ =\  0, \nonumber \\
{U}(1)_X \otimes \left[ {U}(1)_Y \right]^2&\ :&
			x_q - 8 x_u - 2x_d + 3x_\ell - 6x_e &\ =\  0, \nonumber \\
\left[ {U}(1)_X \right]^2 \otimes {U}(1)_Y&\ :&
			{x_q}^2 - {2x_u}^2 + {x_d}^2 - {x_\ell}^2 + {x_e}^2 &\ =\  0, \nonumber \\
\left[ {U}(1)_X \right]^3&\ :&
			{6x_q}^3 - {3x_u}^3 - {3x_d}^3 + {2x_\ell}^3 - {x_\nu}^3 - {x_e}^3 &\ =\  0, \nonumber \\
{U}(1)_X \otimes \left[ {\rm grav.} \right]^2&\ :&
			6x_q - 3x_u - 3x_d + 2x_\ell - x_\nu - x_e &\ =\  0. 
\label{anom-f}
\end{align*}
Hence we obtain the general $U(1)_X$ charge assignment using the Yukawa interaction given in Eq.~\ref{Yukawa} where the charges of the particles can be expressed as a linear combination of the $U(1)_Y$ and $B-L$ charges, as shown in Table~\ref{tab1}.
\section{Relevant vertices}
\label{app:vertices}
In Table.~\ref{tab:cubic-coupling} and~\ref{tab:quartic-coupling}, we list all the relevant cubic and quartic scalar boson couplings which plays role in DM analysis.
\begin{table}[H]
\setlength\tabcolsep{0.25cm}
\centering
\begin{tabular}{| c || c |}
\hline
$\lambda_{abc}$ &  Couplings in terms of Lagrangian parameter \\
\hline
$h_1\chi_R\chi_R$ & $ \lambda_{\Phi\chi}v_{\Phi}\sin(\alpha)+\lambda_{H\chi}v_{H} \cos\alpha + \sqrt{2}\lambda_{\Phi\chi\chi}\sin\alpha $ \\
$h_2\chi_R\chi_R$ & $ \lambda_{\Phi\chi}v_{\Phi}\cos\alpha - \lambda_{H\chi}v_{H}\sin\alpha + \sqrt{2}\lambda_{\Phi\chi\chi}\cos\alpha$ \\
$h_1\chi_I\chi_I$ & $\lambda_{\Phi\chi}v_{\Phi}\sin\alpha+\lambda_{H\chi}v_{H} \cos\alpha-\sqrt{2}\lambda_{\Phi\chi\chi}\sin\alpha$ \\
$h_2\chi_I\chi_I$ & $\lambda_{\Phi\chi}v_{\Phi}\cos\alpha-\lambda_{H\chi}v_{H}\sin\alpha-\sqrt{2}\lambda_{\Phi\chi\chi}\cos\alpha$ \\
\hline
\hline
$Z\chi_I\chi_R$ &   $x_{\Phi}g^{\prime}_1\sin{\theta^\prime}(p^{\mu}_{\chi_R}-p^{\mu}_{\chi_I})$\\
$Z'\chi_I\chi_R$ &  $x_{\Phi}g^{\prime}_1\cos{\theta^\prime}(p^{\mu}_{\chi_R}-p^{\mu}_{\chi_I})$\\
\hline
\hline
$h_1N^3_RN^3_R$   &  $\frac{Y^3_{M}}{\sqrt{2}}\sin{\alpha}$  \\
$h_2N^3_RN^3_R$   &  $\frac{Y^3_{M}}{\sqrt{2}}\cos{\alpha}$ \\
$ZN^3_RN^3_R$   &  $x_{\Phi}g^{\prime}_1\sin{\theta^\prime}\gamma^{\mu}\gamma^{5}$\\
$Z'N^3_RN^3_R$  &  $x_{\Phi}g^{\prime}_1\cos{\theta^\prime}\gamma^{\mu}\gamma^{5}$\\
\hline
\end{tabular}
\caption{The cubic couplings of the DM scalar and fermion.}
\label{tab:cubic-coupling}
\end{table}
\begin{table}[H]
\setlength\tabcolsep{0.25cm}
\centering
\begin{tabular}{| c || c |}
\hline
$\lambda_{abc}$ &  Couplings in terms of Lagrangian parameter \\
\hline
$h_1h_1\chi_{R(I)}\chi_{R(I)}$ & $\lambda_{\Phi\chi}\sin^2\alpha+\lambda_{H\chi}\cos^2\alpha $\\
$h_1h_2\chi_{R(I)}\chi_{R(I)}$ & $\sin\alpha\cos\alpha(\lambda_{\Phi\chi}-\lambda_{H\chi}) $ \\
$h_2h_2\chi_{R(I)}\chi_{R(I)}$ & $\lambda_{\Phi\chi}\cos^2\alpha+\lambda_{H\chi}\sin^2\alpha$\\
\hline
\hline
$ZZ\chi_{R(I)}\chi_{R(I)}$ & $2 x^2_{\Phi}(g^{\prime}_1)^2\sin^2{\theta^\prime} g_{\mu\nu}$\\
$ZZ'\chi_{R(I)}\chi_{R(I)}$ & $2 x^2_{\Phi}(g^{\prime}_1)^2\sin{\theta^\prime}\cos{\theta^\prime} g_{\mu\nu}$\\
$Z'Z'\chi_{R(I)}\chi_{R(I)}$ & $2 x^2_{\Phi}(g^{\prime}_1)^2\cos^2{\theta^\prime} g_{\mu\nu}$\\
\hline
\end{tabular}
\caption{The quartic couplings of the DM scalar.}
\label{tab:quartic-coupling}
\end{table}
\section{Feynman diagrams}
\label{app:feynman diagram}
\begin{figure}[H]
\begin{center}
\includegraphics[height=6cm, width=16cm]{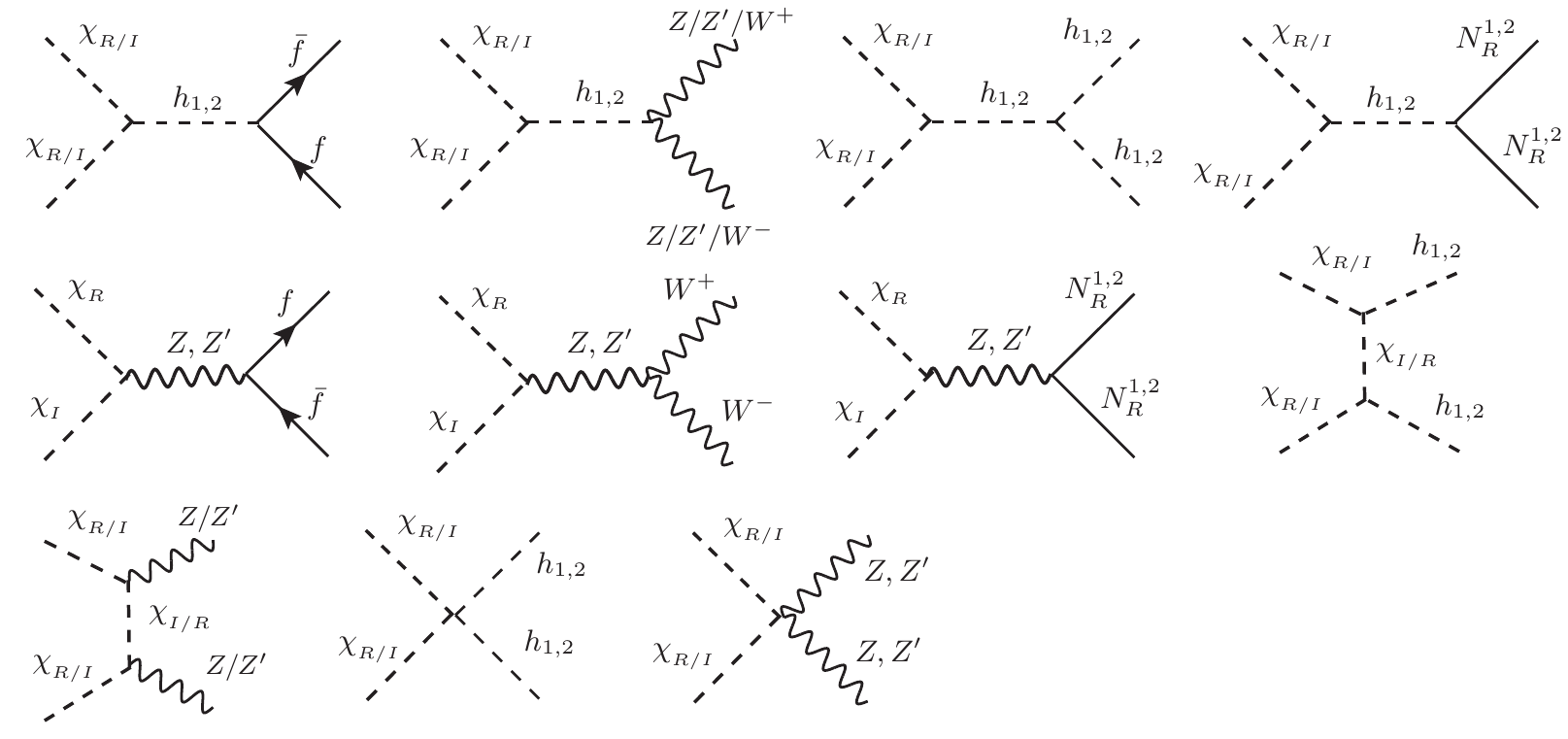}
\caption{Annihilation and coannihilation tree level Feynman diagrams contributing to the relic abundance of $\chi_R$.}
\label{fig:annihilation-diagram-scalar}
\end{center}
\end{figure}

\begin{figure}[H]
\begin{center}
\includegraphics[height=6cm, width=16cm]{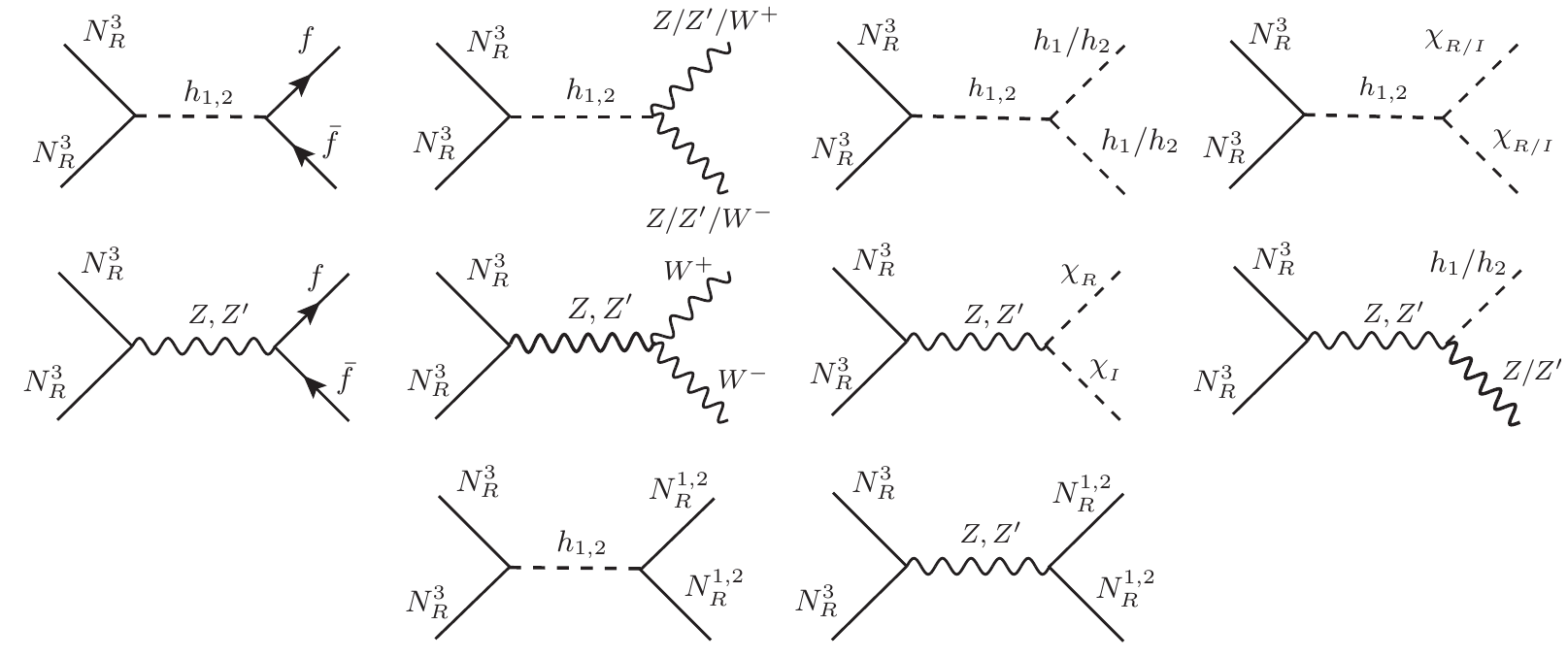}
\caption{Annihilation and coannihilation tree level Feynman diagrams contributing to the relic abundance of $N_R^3$.}
\label{fig:annihilation-diagram-fermion}
\end{center}
\end{figure}

\begin{figure}[H]
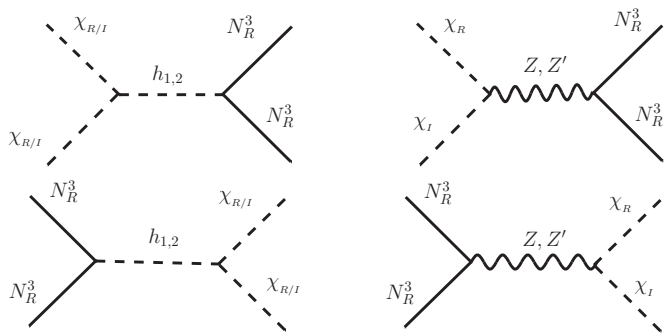

\begin{center}
\includegraphics[width=0.5\textwidth]{feyn_diag_U1XINR.pdf}\\
\includegraphics[width=0.5\textwidth]{feyn_diag_U1XnuI.pdf}
\caption{The Feynman diagrams that contribute to the process of DM conversion, $\chi\chi\leftrightarrow N_3 N_3$.}
\label{fig:annihilation-diagram-conversion}
\end{center}
\end{figure}

The relevant Feynman diagram for relic density analysis of scalar and fermion DM are shown in Figs.~\ref{fig:annihilation-diagram-scalar}, \ref{fig:annihilation-diagram-fermion} and \ref{fig:annihilation-diagram-conversion}, respectively.
\bibliographystyle{utphys}
\bibliography{bibitem}
\end{document}